\documentclass[12pt]{article}
\usepackage{latexsym,epsfig,graphicx,amsmath,amssymb,amscd,undertilde,multirow,chicago,psfrag,paralist,dsfont}

\newcommand{\thetavec}{{\boldsymbol{\theta}}}

\newcommand{\Sigmavec}{{\boldsymbol{\Sigma}}}

\newcommand{\yvec}{{\boldsymbol{y}}}
\newcommand{\zvec}{{\boldsymbol{z}}}

\newcommand{\betavec}{{\boldsymbol{\beta}}}

\newcommand{\NOR}{{\rm N}}
\newcommand{\ER}{{\rm ER}}

\newcommand{\wh}{\widehat}

\newcommand{\Xvec}{\boldsymbol{X}}

\newcommand{\xvec}{\boldsymbol{x}}

\newcommand{\D}{\mathcal{D}}
\newcommand{\calS}{\mathcal{S}}
\newcommand{\yhat}{\widehat{y}}

\begin{document}


\title{Prediction of High-Performance Computing Input/Output Variability and Its Application to Optimization for System Configurations}

\author{\small Li Xu$^1$, Thomas Lux$^2$, Tyler Chang$^2$, Bo Li$^2$, Yili Hong$^1$, Layne Watson$^2$,\\
\small Ali Butt$^2$, Danfeng Yao$^2$, and Kirk Cameron$^2$\\
\small $^1$Department of Statistics, Virginia Tech, Blacksburg, VA 24061\\
\small $^2$Department of Computer Science, Virginia Tech, Blacksburg, VA 24061\\
}

\date{}

\maketitle
\begin{abstract}
Performance variability is an important measure for a reliable high performance computing (HPC) system. Performance variability is affected by complicated interactions between numerous factors, such as CPU frequency, the number of input/output (IO) threads, and the IO scheduler. In this paper, we focus on HPC IO variability. The prediction of HPC variability is a challenging problem in the engineering of HPC systems and there is little statistical work on this problem to date. Although there are many methods available in the computer experiment literature, the applicability of existing methods to HPC performance variability needs investigation, especially, when the objective is to predict performance variability both in interpolation and extrapolation settings.  A data analytic framework is developed to model data collected from large-scale experiments. Various promising methods are used to build predictive models for the variability of HPC systems.  We evaluate the performance of the methods by measuring prediction accuracy at previously unseen system configurations. We also discuss a methodology for optimizing system configurations that uses the estimated variability map. The findings from method comparisons and developed tool sets in this paper yield new insights into existing statistical methods and can be beneficial for the practice of HPC variability management. This paper has supplementary materials online.


\end{abstract}


\newpage

\section{Introduction}
\subsection{The Problem}
High performance computing (HPC) commonly refers to the aggregation of computing power to obtain much higher performance than a typical desktop computer or workstation. HPC is widely used to solve large-scale problems in various areas such as science, engineering, and business. While improving the performance of HPC systems attracts lots of research, managing the performance variability of HPC systems is also an important dimension of HPC system management that can not be ignored. A common manifestation of performance variability is the variation from run to run in the execution time for a particular task.

\begin{figure}
\begin{center}
\includegraphics[width=.5\textwidth]{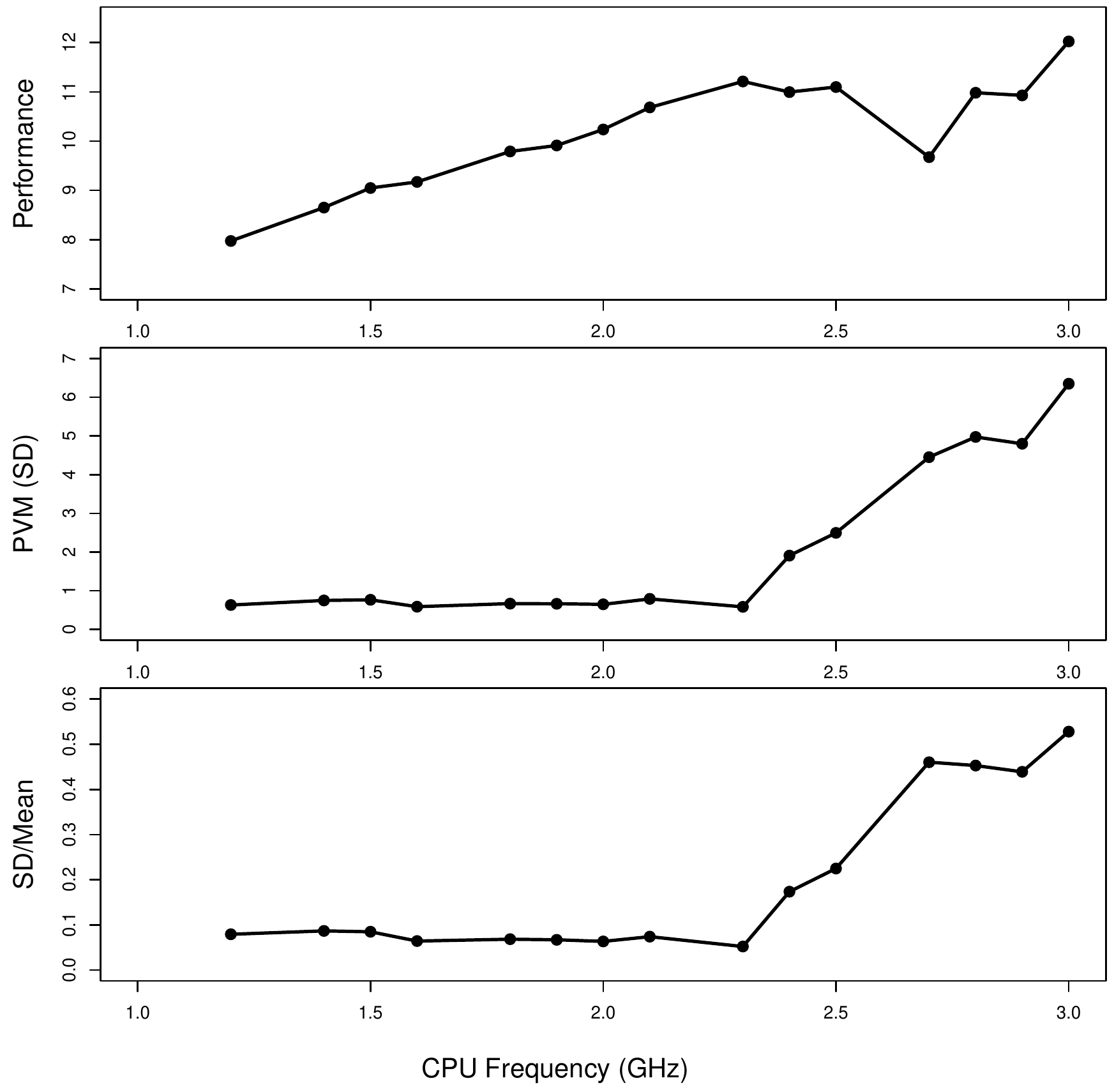}
\end{center}
\caption{Example of performance variability of IO throughput (in units of KB/s$\times10^{7}$) as a function of CPU frequency, while keeping other variables fixed. The top panel shows the mean throughput, which is the performance measure. The middle panel shows the standard deviation (SD), which is used as the performance variability measure (PVM). The bottom panel shows the ratio of the SD over the mean, which is referred to as the coefficient of determination in statistics.}\label{fig:pvm.illustration}
\end{figure}

In a more specific example, the middle panel of Figure~\ref{fig:pvm.illustration} shows the performance variability of input/output~(IO) throughput (data transfer speed, in units of $10^{7}$ KB/s) as a function of CPU frequency with other system parameters fixed. Following \shortciteN{Cameron-MOANA-2019}, we use the standard deviation of IO throughput from multiple runs as a performance variability measure (PVM). While deferring the details of Figure~\ref{fig:pvm.illustration} to later sections, Figure~\ref{fig:pvm.illustration} shows that the performance variability increases as the CPU frequency increases. Although one can generally obtain higher throughput when increasing the CPU frequency as shown in the top panel of Figure~\ref{fig:pvm.illustration}, the variability of the throughput from different runs also increases when the CPU frequency increases. For high CPU frequencies, the variability is high and the standard deviation can exceed 50\% of the mean performance as shown by the bottom panel of Figure~\ref{fig:pvm.illustration}, which plots the ratio of the standard deviation over the mean throughput. Thus, high performance variability can be a hindrance for system development because it makes the evaluation of a system challenging when the performance is inconsistent across individual runs. There are also other factors such as imbalanced tasks caused by variability that can affect overall system performance and efficiency.

In the big picture, performance variability in HPC systems is common and critical. Exponential increases in complexity and scale make variability a growing threat to sustaining HPC performance at exascale. However, the performance variability is affected by complicated interactions of numerous factors, such as CPU frequency, the number of threads, and the IO scheduler. The study of performance variability is an emerging area in computer science. The fundamental questions are: how do system variables affect the performance variability, can we predict performance variability for new system design/configurations, and how can we manage performance variability?

To answer these questions, computer scientists have completed large scale experiments to collect the necessary data. In this paper, we present the statistical modeling of HPC IO variability. The prediction and management of variability is a challenging problem in the engineering of HPC systems and there is little statistical work on this problem. Specifically, the objective is to conduct statistical modeling and make predictions about performance variability. In particular, we use various statistical methods to describe system variability with system configuration variables as inputs, which we refer to as a variability map construction. With the variability map, we make predictions for new design points (i.e., new system configurations). We examine two types of predictions: interpolation and extrapolation, both of which are of interest to computer scientists. Using prediction accuracy as the criterion, we evaluate and compare the performance of different methods. This comparison provides insights into the effectiveness of different methods for predicting HPC performance variability. While it is impossible to eliminate variability, the statistical models and predictions can help manage the variability. Being able to predict variability allows one to identify system configurations that can lead to large variability. Thus, one can avoid these system configurations and design a system in which performance and variability are both optimized.

\subsection{Related Literature and the Contribution of This Work}
In the area of statistics, computer experiments are constructed to emulate a physical system. Because these are meant to replicate some aspects of a system in detail, they often do not yield an analytic solution. Due to the complexity and expense of evaluating system behavior, a surrogate model is usually used to describe the system behavior under some configuration. A surrogate model is an engineering method used when an outcome of interest can not be easily or directly measured, so a model of the outcome is used instead. Popular surrogate models (\shortciteNP{compexpsummary}) include response surface models, Kriging, gradient-enhanced Kriging, support vector machines, space mapping, and artificial neural networks (\shortciteNP{surrogatemodel}). In our work, we mainly focus on the following surrogate models: response surface models, Gaussian process based methods, inverse distance weighting methods, and non-parametric regression models. A good reference book on response surface models is \citeN{MyersMontgomeryAnderson-Cook2016}.

The Gaussian process model is a popular tool in computer experiments (e.g., \citeNP{santner2010design}). However, a Gaussian process only forms a stationary surface. When the underlying surface has steep behavior, the Gaussian process may not be able to capture those steep changes. Several methods have been proposed to extend the Gaussian process to form a non-stationary surface. \citeN{Chipman2002} proposed a Bayesian treed Gaussian process that uses a reversible-jump Monte Carlo Markov chain (MCMC) to build the tree structure. \citeN{tgpjasa} built a treed Gaussian process and \citeN{dynatree} developed dynamic trees. Tree-based non-stationary Gaussian processes in \citeN{tgpjasa} and \citeN{dynatree} used similar priors to build tree structures, while the dynamic tree's prior extended the method in \citeN{CGM98}. \citeN{bartanas} proposed an additive Gaussian process. In addition, \citeN{QianWuWu2008} proposed a framework for building Gaussian process models that can handle qualitative and quantitative factors.  \citeN{ZhangNotz2014} provided a comprehensive review of computer experiments with qualitative and quantitative variables. \citeN{DengHungLin2015} introduced the marginally coupled designs for computer experiments that can consider both qualitative and quantitative variables.

Spline bases are widely used when the relationship between input and response is highly nonlinear. In particular, the multivariate adaptive regression splines (MARS) in \citeN{friedman1991} is a technique that uses an iterative approach to select spline bases for the approximation of an unknown surface. The criterion used in selecting the bases is related to the entire data. MARS only assumes a polynomial relationship between the response and the input variables, and can capture very complex relationships adaptively. A generalized cross validation procedure is used to determine the best model. More details on the generalized cross validation can be found in \citeN{statisticallearning}. Adaptive splines are also adopted in other approximations to improve accuracy, for example, in neural networks (e.g., \shortciteNP{splineNN}), and Gaussian process modeling (e.g., \shortciteNP{RSSB}). \citeN{BenAriSteinberg2007} compared Kriging with MARS and projection pursuit regression in the modeling of data from computer experiments.

For inverse distance weighting methods, the assigned values to unknown points are calculated with a weighted average of the values available at the known points. The earliest inverse distance weighting method is Shepard's method (\citeNP{Shepard}). Many extensions have been made based on Shepard's work (e.g., \citeNP{Gordon:1978}, \citeNP{Renka1}, and \citeNP{Berry:1999}).

Although there are many methods available in the computer experiment literature, the applicability of existing methods to HPC performance variability is not clear. We investigate the suitability of several methods in the literature and provide new insights into the practical application of those methods in HPC variability management. The findings of method comparisons and developed tool sets in this paper can be beneficial for the practice of HPC variability management and system building.

\subsection{Overview}
The rest of the paper is organized as follows. Section~\ref{sec:data.collection} introduces the experiment setup and describes the data collection process. Section~\ref{sec:var.map.constr} introduces various methods that can be used for variability map construction. Section~\ref{sec:pred.meth.comp} compares different methods for their performance in interpolation and extrapolation. Section~\ref{sec:optim.per.var} presents the optimization of performance variability. Section~\ref{sec:conclusion} contains some concluding remarks and areas for future research.

\section{Experiment Setup and Data Collection}\label{sec:data.collection}
\subsection{Performance Variability Measure}
While HPC performance can be a general term, we focus on one specific metric, which is the input/output (IO) throughput to persistent memory (i.e., hard disk drives and solid state drives). The modeling and analysis approach used in this paper, however, can be extended to other metrics. Here, the IO throughput is defined as the data transfer speed in terms of kilobytes per second (KB/s). Even under the same system (i.e., hardware and software) configuration, different runs of the same task will end up with different IO throughput. That is, performance variability exists in IO throughput. Specifically, the performance variability measure (PVM) used in this paper is the standard deviation of the IO throughput from multiple runs of the same task. Here, runs refer to the replicates of IOzone experiments under a given system configuration. For each run, we run the IOzone benchmark under a fixed configuration to obtain an IO throughput measurement. This process is repeated to obtain multiple measurements. We use the standard deviation of those throughput measurements as the PVM. The objective is to study how system variables affect the IO throughput, through statistical modeling of the relationship between system variables (including hardware and software configurations) and the performance variability.

\subsection{Experiment Setup}
The IO throughput to persistent memory is affected by three categories of variables: hardware configurations, operating system configurations, and application configurations. To study how the system configurations affect IO performance variability, complicated large-scale experiments were conducted by computer scientists. In this section, we introduce the setup of these experiments. The response variable is the throughput variability tested by running the IOzone benchmark on a server under the same system configuration  $\xvec$, with a relatively large number of repeated runs. In particular, 40 runs were used under budget constraints. The server is configured with a dedicated 2TB HDD on a 2 socket/4 core Intel Xeon E5-2623 v3 (Haswell) platform. The memory is 32 GB DDR4. The number of hyperthreads is 2 per core.  The Linux operating system used is Ubuntu 14.04. More details can be found in \shortciteN{Cameron-MOANA-2019}.

For system configurations, seven important variables were selected by computer scientists, which consist of both categorical and continuous variables. Table~\ref{tab:exp.setup} shows the summary information for the seven variables that were used in the IOzone throughput experiments. On the hardware configurations, changing CPU clock frequency is an important method for adjusting system performance and energy efficiency. On the experiment server, there are 15 levels of frequencies ranging from 1.2 to 3.0 GHz. On the operating system configuration, one can change the IO scheduling policy of the host system as well as the hypervisor by which virtual systems are controlled. Both the host system and virtual machine (VM) have 3 levels of IO schedulers, which are CFQ, DEAD, and NOOP.

The application configuration has four variables, which are the IO Operation Mode, the Number of Threads, File Size (KB), and Record Size (KB). The IO Operation Mode has 13 levels, which cover almost all the common IO operations on HPC systems. For example, Fread and Fwrite represent different types of reading and writing tasks, respectively. The Number of Threads has 9 levels, ranging from 1 to 256 by powers of 2. The values of File Size and Record Size are chosen in pairs as there are some constraints on the selection of values. In particular, the File Size has to be larger than the Record Size, and has to be a multiple of the Record Size. For example, the sixth row of Table~\ref{tab:exp.setup} shows the valid combinations of File Size and Record Size used in the training set.

\begin{table}
\begin{center}
\caption{Summary information for variables that are used in the IOzone throughput experiments to collect the training set and extrapolation test set. Note in the File Size by Record Size row, the first number is File size and the second number is Record Size. The number of all the combinations in this table is $13\times3\times3\times15\times9\times6=94770$. The settings, at which the Frequency and the Number of Threads are (2.8, 256), (2.9, 256), (3.0, 256), (3.0, 64), and (3.0, 128) given the File Size and Record Size is (256, 32),  are set aside for extrapolation test set, which excludes $13\times3\times13\times5\times1=585$ settings with large values on the Frequency and the Number of Threads. Thus, the number of points in the training set is $94770-585=94185$.}\label{tab:exp.setup}
\vspace{.5em}
\begin{tabular}{c|c|c|c}\hline\hline
\multirow{2}{*}{Category}      & \multirow{2}{*}{Variable} & No. of  & \multirow{2}{*}{Values}\\
                               &                           & levels  &                        \\\hline
Hardware      & CPU Clock Frequency       & \multirow{2}{*}{15} & 1.2, 1.4, 1.5, 1.6, 1.8, 1.9, 2.0, 2.1 \\
              &    (GHz)                        &    & 2.3, 2.4, 2.5, 2.7, 2.8, 2.9, 3.0\\\hline
Operating     & IO Scheduler         & 3  & CFQ, DEAD, NOOP \\\cline{2-4}
System        & VM IO Scheduler        & 3  & CFQ, DEAD, NOOP\\\hline
\multirow{7}{*}{Application}& \multirow{4}{*}{IO Operation Mode}  & \multirow{4}{*}{13} & {\small Fread, Pread, Re-read, Randomread}\\
              &                                 &    & {\small Read, ReverseRead, Strideread }    \\
              &                                 &    & {\small Fwrite, Pwrite, Randomwrite, Rewrite}    \\
              &                                 &    & {\small Initialwrite, Mixedworkload}\\\cline{2-4}
              & Number of Threads & 9  & 1,   2,   4,   8,  16,  32,  64, 128, 256\\\cline{2-4}
              & File Size (KB) by               & \multirow{2}{*}{6}  & (64, 32), (256, 32), (256, 128)\\
              & Record Size (KB)                &    & (1024, 32), (1024, 128), (1024, 512)\\\hline\hline
\end{tabular}
\end{center}
\end{table}

\subsection{Prediction Problems and Data Collection}
Two kinds of predictions arise in the setting of HPC variability management, which are interpolation and extrapolation. To test the predictability of different methods, we collect three sets of data: the training set $(\calS_t)$, the interpolation test set $(\calS_i)$, and the extrapolation test set $(\calS_e)$.  Let $\xvec$ be the vector representing the system variables (i.e., system parameters). We build a prediction model based on the training set $\calS_t$, and we then use the built prediction model to make a prediction for the PVM under a new configuration $\xvec_w$. A prediction problem is called an \emph{interpolation} if $\xvec_w$ lies within the convex hull of $\calS_t$, and a prediction problem is called an \emph{extrapolation} if $\xvec_w$ falls outside the convex hull of $\calS_t$.

Regarding $\calS_t$, $\calS_i$, and $\calS_e$, each dataset contains eight columns with one column for the PVM, which is the response/output variable, and seven columns for the explanatory/input variables. The IO Scheduler, VM IO Scheduler, and IO Operation Mode are treated as categorical variables, and the Frequency, Number of Threads, File Size and Record Size are treated as continuous variables.

\begin{figure}
\begin{center}
\begin{tabular}{cc}
\includegraphics[width=.47\textwidth]{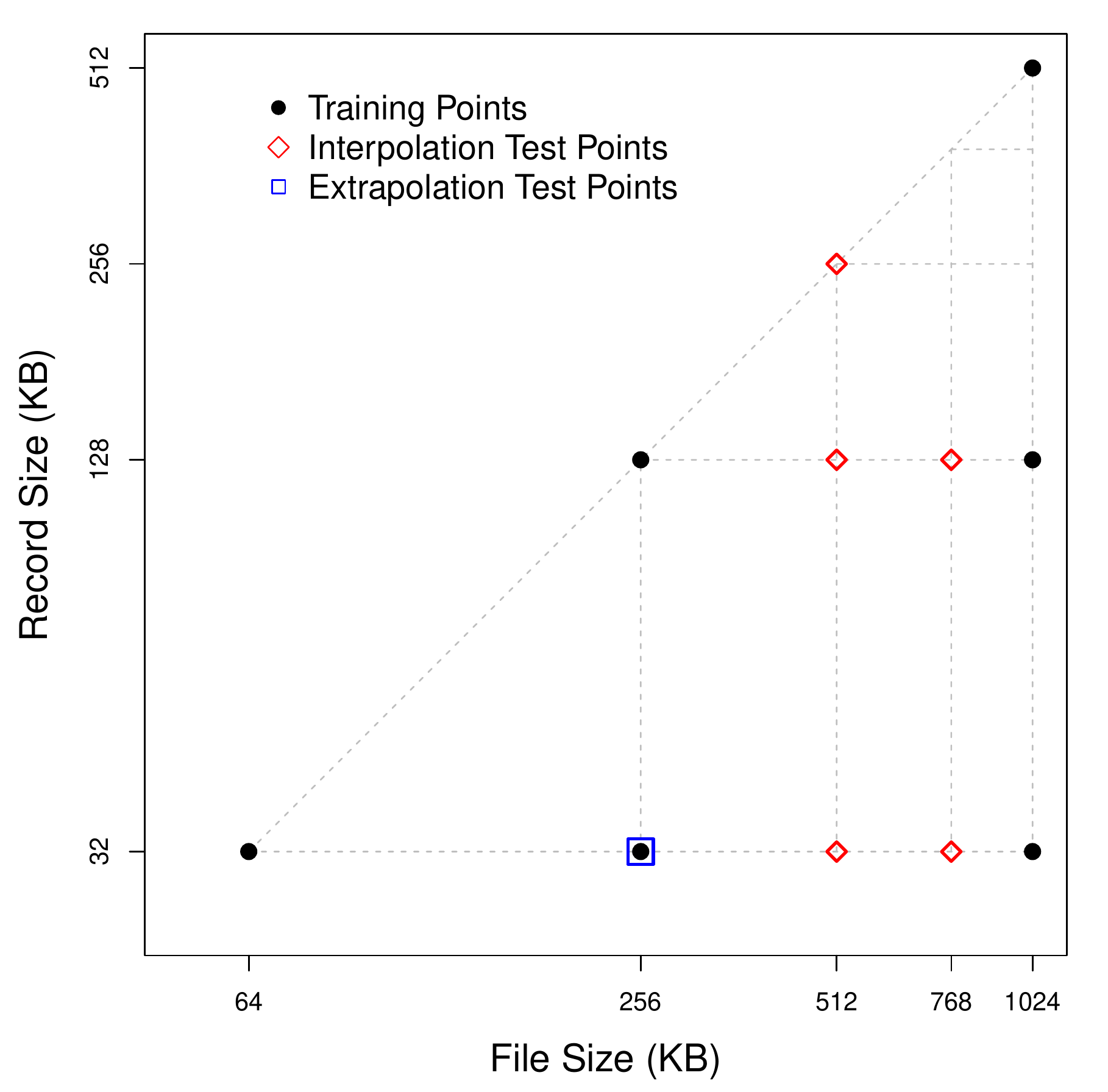}&
\includegraphics[width=.45\textwidth, angle=90]{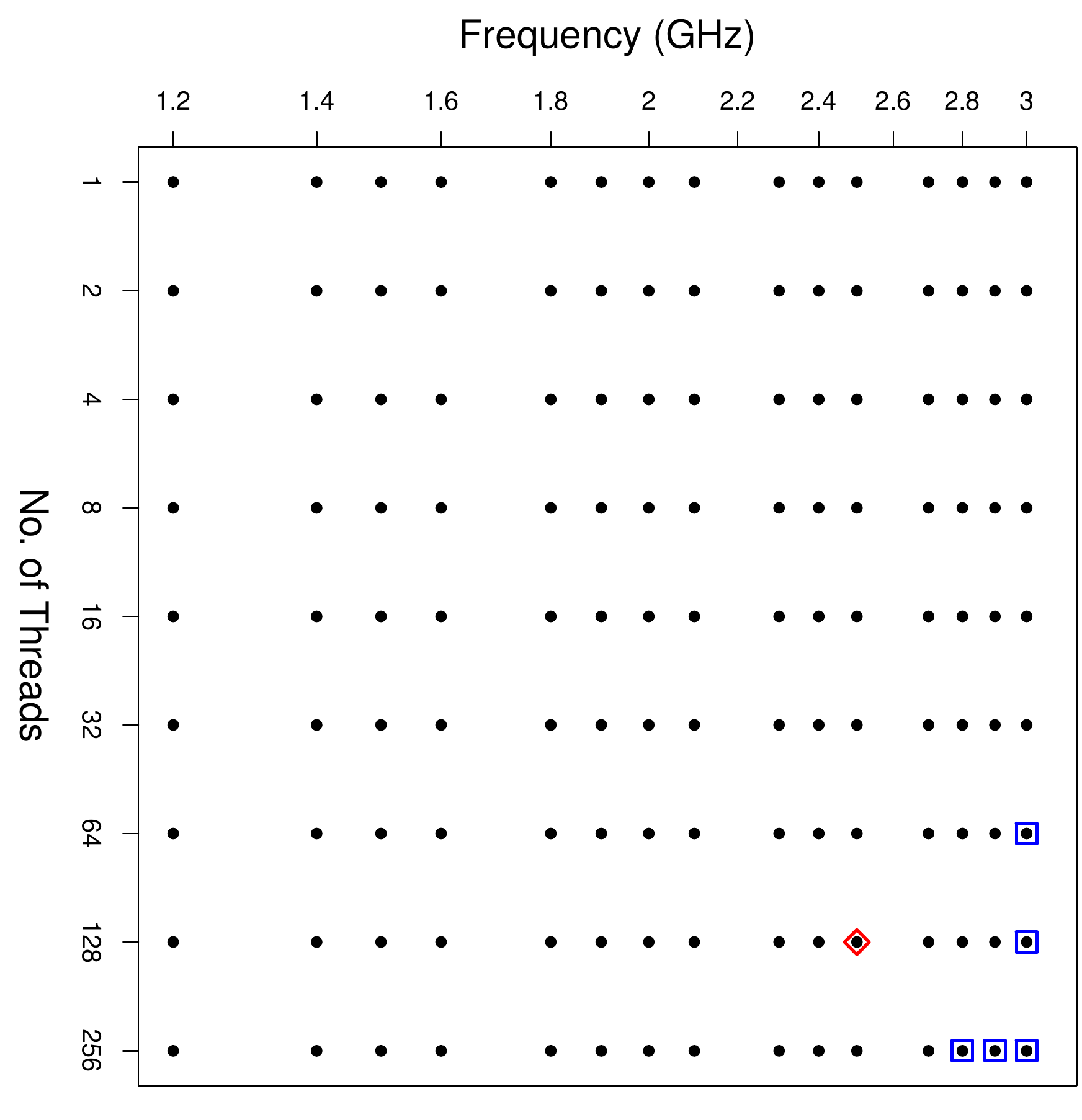}\\
(a) File Size and Record Size & (b) Number of Threads and Frequency
\end{tabular}
\end{center}
\caption{Illustration of design points for File Size, Record Size, Number of Threads, and Frequency. The points marked by black dots are for the training set. The points marked by black dots with a square are for the extrapolation test set, and the points marked by diamonds are for the interpolation test set.}\label{fig:design.points}
\end{figure}

The training set, $\calS_t$, is used to build a model for predicting variability. As explained in the caption of Table~\ref{tab:exp.setup}, the training set consists of
\begin{align}\label{eqn:training.points}
13\times3\times3\times15\times9\times6-13\times3\times3\times5\times1=94185
\end{align}
points. The training set consists of all combinations of the variables (i.e., $13\times3\times3\times15\times9\times6=94770$) with $13\times3\times3\times5\times1=585$ points excluded for the extrapolation test set. Figure~\ref{fig:design.points} illustrates the design points for File Size, Record Size, Number of Threads, and Frequency, as shown by black dots.

The subtracted points in \eqref{eqn:training.points} are those at which the Frequency and the Number of Threads are (2.8, 256), (2.9, 256), (3.0, 256), (3.0, 64), and (3.0, 128) given the File Size and Record Size is (256, 32), which are used for extrapolation testing. Figure~\ref{fig:design.points} displays the points that are used for extrapolation testing purposes, which are marked by squares. Note that the points are on the right upper corner of the combination of Frequency and the Number of Threads, representing large values of the two variables. Thus, the main purpose is to test the extrapolation of the Frequency and the Number of Threads to large values.

There are an additional $3\times3\times13\times5\times1\times1=585$ points used for interpolation testing purposes. Those points are marked by diamonds in Figure~\ref{fig:design.points}. Specifically, the chosen points are the File Size and Record Size at (512, 32), (512, 128), (512, 256) (768,  32), and (768, 128), given the Number of Threads is 128 and the Frequency is 2.5~GHz. The points lie in the interior of or on the convex hull of File Size and Record Size points. Thus, the main purpose is to test the interpolation of the File Size and Record Size.

Table~\ref{tab:test.set.setup} shows the summary of the experimental setup for the collections of the two test sets, $\calS_i$ and $\calS_e$. In our data collection, we have more points on the Frequency and Number of Threads combinations than that of the File Size and Record Size combination. Because extrapolation is usually more challenging than interpolation, we chose to test the ability of extrapolation on the Frequency and Number of Threads combinations. For interpolation, we chose to test the ability on the File Size and Record Size combination. Although there were only six combinations for the File Size and Record Size combination, interpolation can still be made based on this information provided by the data.

\begin{table}
\begin{center}
\caption{Summary of the experimental setup for the collections of the two test sets.}\label{tab:test.set.setup}
\vspace{.5em}
\begin{tabular}{c|c|c}\hline\hline
\multirow{2}{*}{Variable} & Interpolation  & Extrapolation\\
                          & Test Set $\calS_i$      & Test Set $\calS_e$    \\\hline
IO Scheduler             & All 3 levels   & All 3 levels \\\hline
VM IO Scheduler          & All 3 levels   & All 3 levels \\\hline
IO Operation Mode        & All 13 levels  & All 13 levels\\\hline
File Size by              & (512,  32), (512, 128), (512, 256)  &  256         \\\cline{1-1}\cline{3-3}
Record Size               & (768,  32), (768, 128)              &    32\\\hline
Frequency                 & 2.5    & (2.8, 256), (2.9, 256), (3.0, 256)     \\\cline{1-2}
Number of Threads         & 128    & (3.0, 64), (3.0, 128)    \\\hline\hline
Total No. of Points & 585  &  585 \\\hline\hline
\end{tabular}
\end{center}

\end{table}

\section{Variability Map Construction}\label{sec:var.map.constr}
\subsection{Data Visualization and Variability Map}
\begin{figure}
\begin{center}
\begin{tabular}{cc}
\includegraphics[width=.45\textwidth]{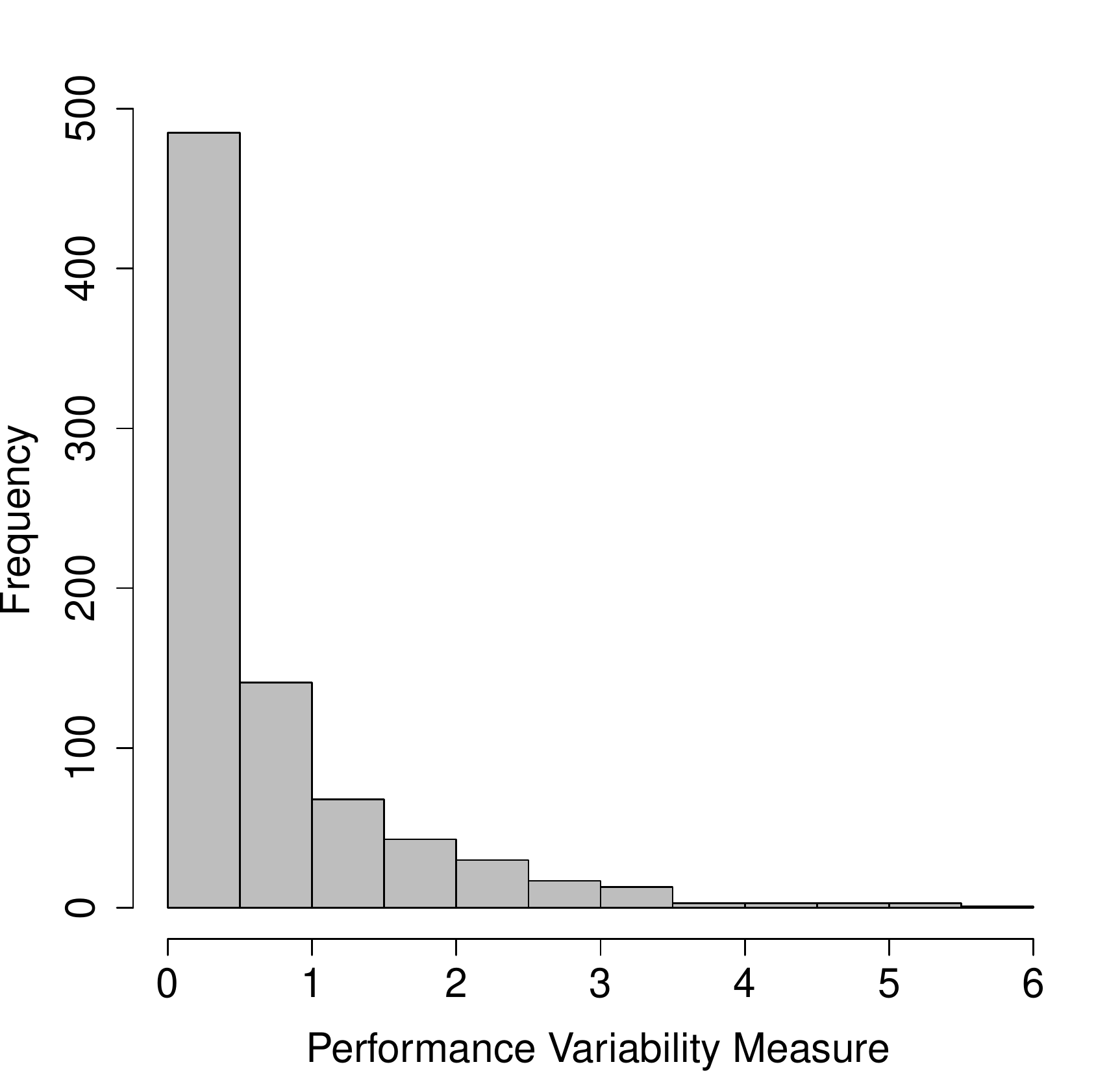}&
\includegraphics[width=.45\textwidth]{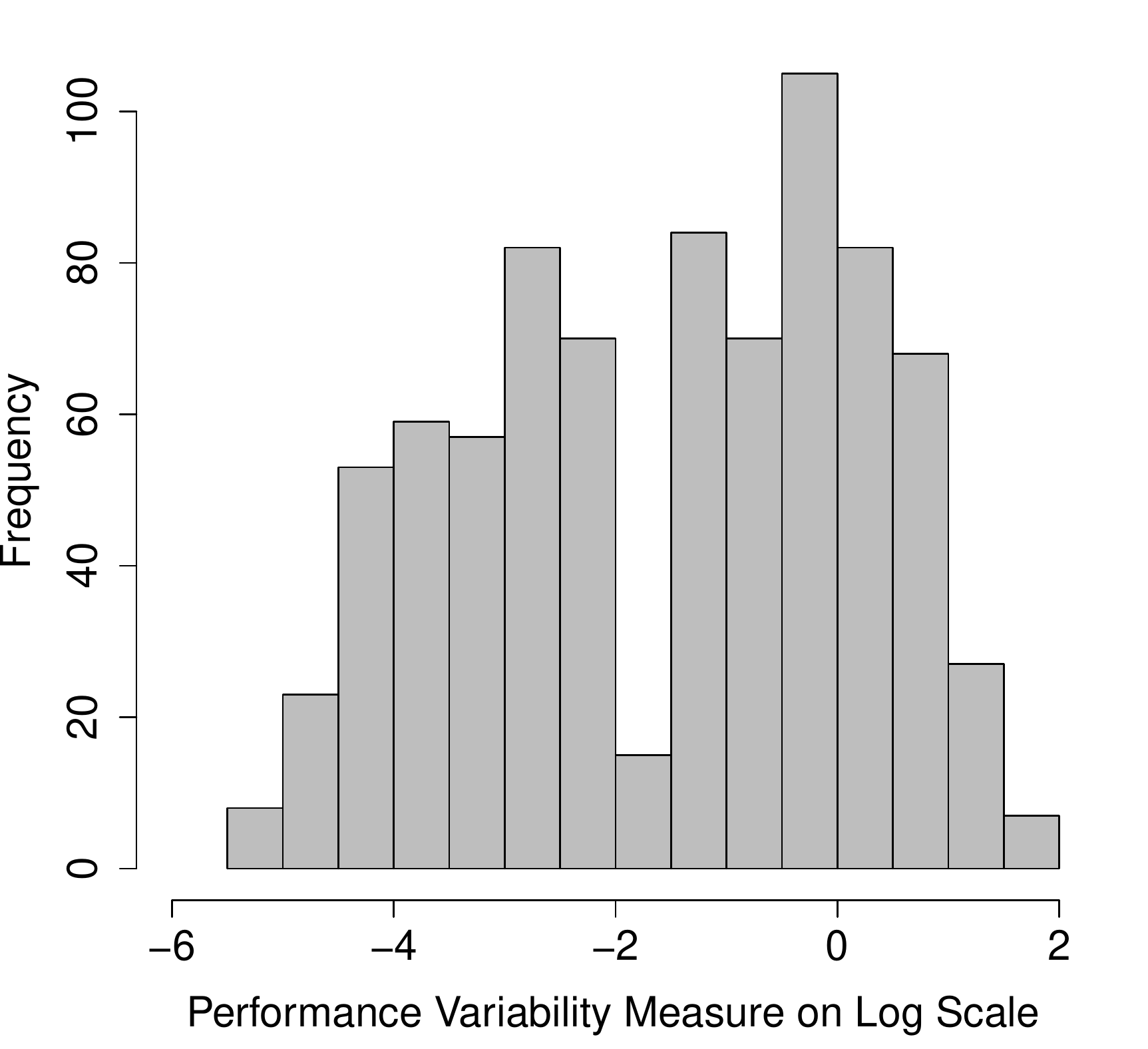}\\
(a) PVM& (b) Log of PVM
\end{tabular}
\end{center}
\caption{One example of a set of system variability before and after log transformation (in a scale of $10^7$). The IO Operation Mode is Fwrite, IO Scheduler is CFQ, and the VM IO Scheduler is NOOP, with all combinations of File Size and Record size.}\label{fig:hist}
\end{figure}

In this section, we first do some data visualization to explore the data. Figure~\ref{fig:hist} shows one example of a set of system variables before and after log transformation using the training set. Note that the scale of the IO throughput is in the magnitude of $10^7$. Because a variable and its standard deviation have the same scale, the PVM (i.e., the standard deviation of IO throughputs) is also in the magnitude of $10^7$. The system configuration for the dataset shown in the figure is chosen as follows: the IO Operation Mode is Fwrite, IO Scheduler is CFQ, and the VM IO Scheduler is NOOP. These categorical configurations were chosen by computer scientists as the most likely selection to be used in practice. With this configuration, Figure~\ref{fig:hist} shows that the PVM of the 805 points (i.e., configurations) is right skewed. After the log transformation, the PVM of the 805 points shows a bimodal behavior. Figure~\ref{fig:hist}(b) also allows us to see that for certain configurations, the variability tends to be small, while for other configurations, the variability tends to be large.

\begin{figure}
\begin{center}
\begin{tabular}{cc}
\includegraphics[width=.45\textwidth]{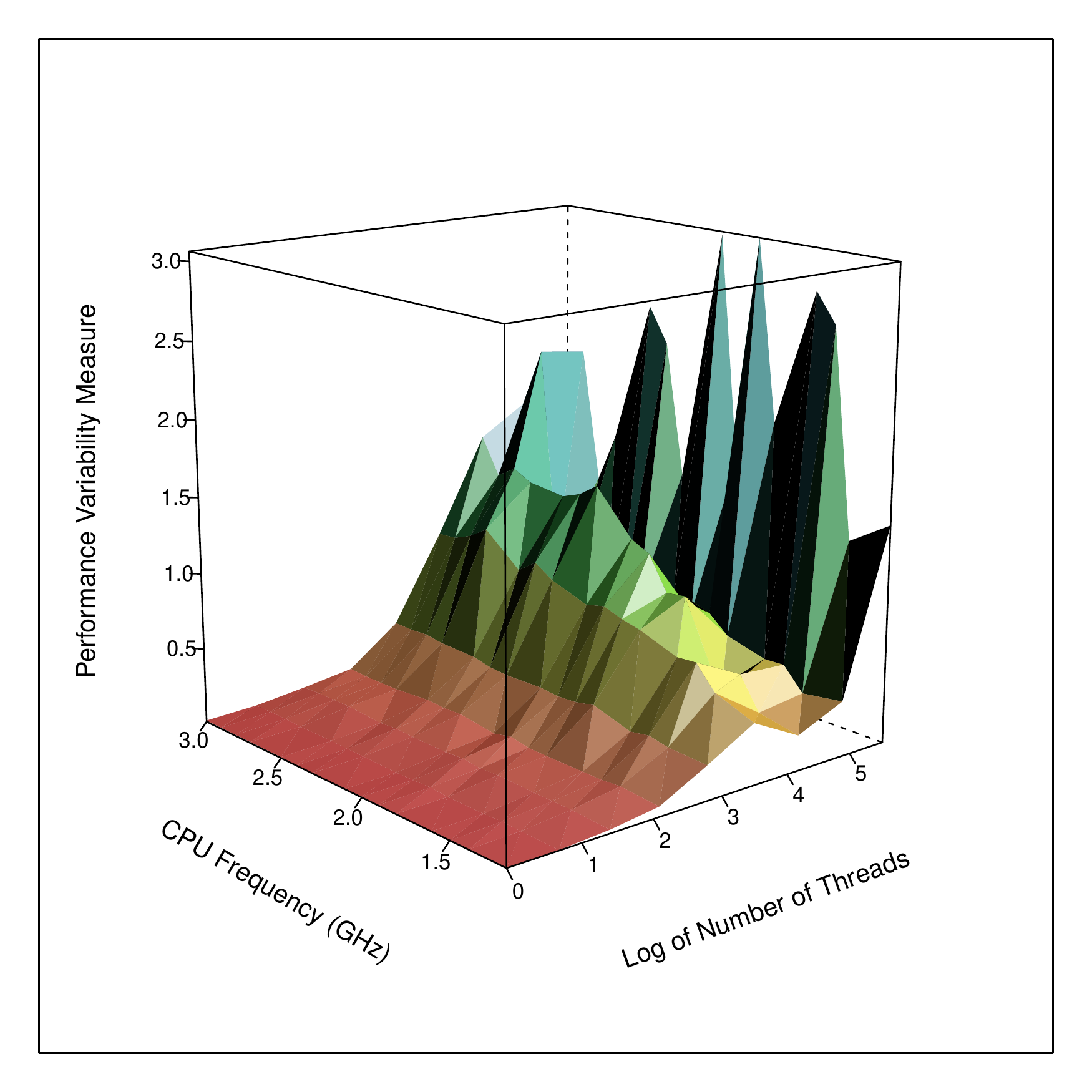}&
\includegraphics[width=.45\textwidth, angle=90]{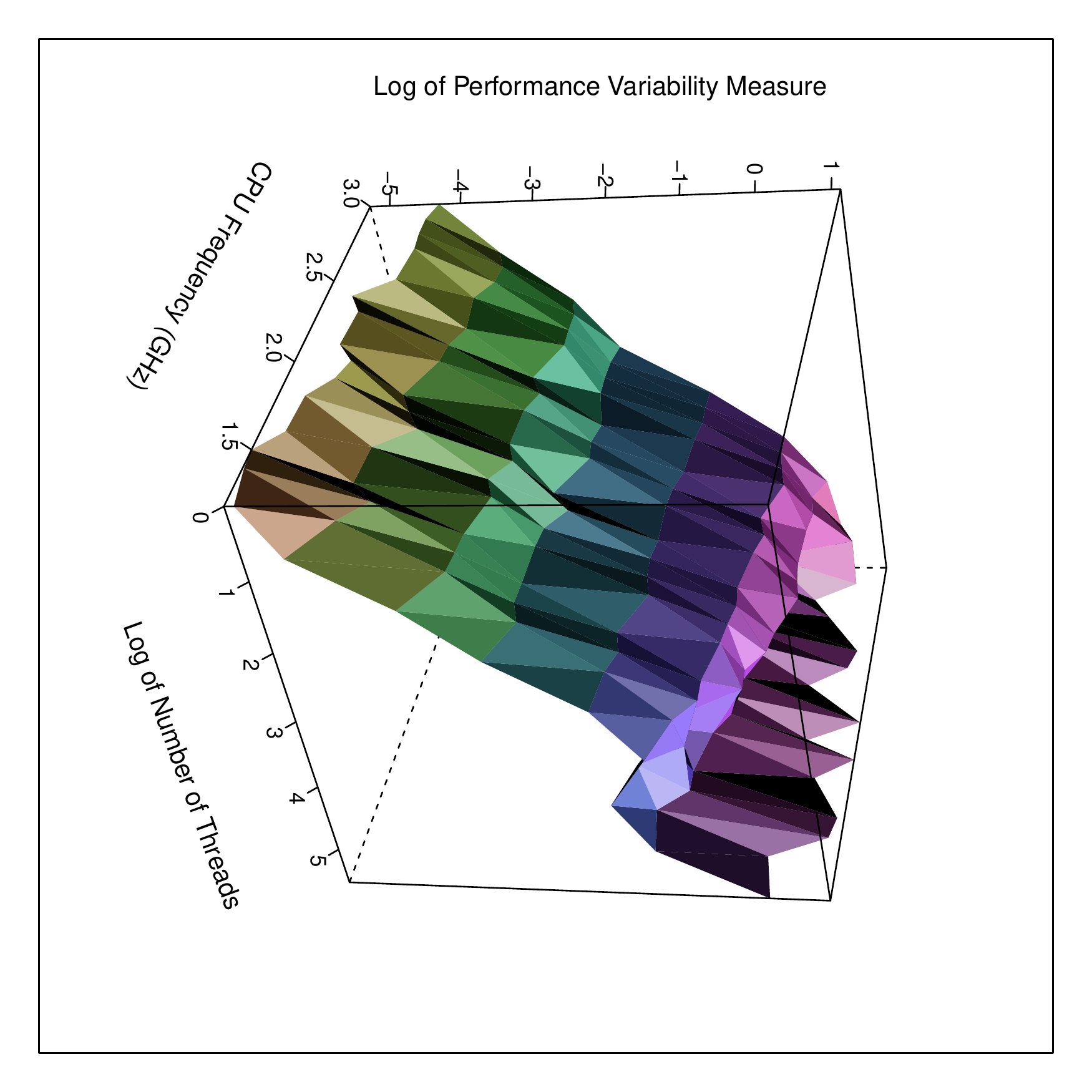}\\
(a) PVM& (b) Log of PVM
\end{tabular}
\end{center}
\caption{Perspective plots of PVM as a function of CPU Frequency and the Number of Threads, before and after taking a log transformation. The IO Operation Mode is Fwrite, IO Scheduler is CFQ, the VM IO Scheduler is NOOP, the File Size is 1024, and the Record Size is 512.}\label{fig:pesp}
\end{figure}

To further visualize how the PVM changes across different system configurations, Figure~\ref{fig:pesp} shows the perspective plots of the PVM as a function of CPU Frequency and the Number of Threads, before and after taking a log transformation. The IO Operation Mode is Fwrite, IO Scheduler is CFQ, the VM IO Scheduler is NOOP, the File Size is 1024, and the Record Size is 512. The figure shows that, in general, the PVM increases as the Frequency and the Number of Threads increase. However, the exact relationship is complicated, and cannot be described by simple functional relationships. While the figures can only show the pattern for two variables, the actual relationship is more complicated due to many other variables that affect the PVM. Note that Figure~\ref{fig:pesp} only shows the pattern for one slice of the data. The shapes of the perspective plots could differ for different configurations.

We introduce the concept of the variability map to describe the functional relationship between the system configuration and the PVM. The variability map is defined as a function $f(\xvec)$ that gives the PVM at $\xvec$. With the variability map $f(\xvec)$ estimated from the training dataset, one can use it to characterize and predict variability for a given configuration $\xvec$.

We focus on the modeling of continuous variables in this paper. For categorical variables, we do a separate estimation for the variability map for each unique combination of the variables for all candidate prediction methods, except for the categorial Gaussian process (CGP) model. In particular, the IO operation Mode, IO Scheduler, and VM IO Scheduler have 117 unique combinations for their levels. Thus, 117 separate variability maps will be constructed. For certain prediction techniques, such as the linear Shepard (LSP) method, they are not designed for categorical variables. Thus, we need to handle each level of categorical variables separately. This treatment, however, does not limit their practical use because one has to specify the level of the categorical variable in the prediction. That is, we can still use the separate variability maps to generate the prediction, given the level of the categorical variable. The CGP model is able to handle categorical variables to some extent.

Without loss of generality, we only describe the construction of one particular variability map. We define $\xvec_i$ as follows,
\begin{align}\label{eqn:x.def}\nonumber
\xvec_i=&(\textrm{Log of File Size},\, \textrm{Log of Record Size},\, \textrm{Log of No. of Threads},\, \textrm{CPU Frequency})'\\
=&(x_{i1}, x_{i2}, x_{i3}, x_{i4})'.
\end{align}
The training set is denoted by $\{y_i, \xvec_i\},\, i=1,\cdots, n$, where $y_i$ is the PVM under the system configuration $\xvec_i=(x_{i1}, x_{i2}, x_{i3}, x_{i4})'$ for a particular combination of the IO operation Mode, IO Scheduler, and VM IO Scheduler. Here $n=805$ is the number of data points for each variability map.

\subsection{Description of Candidate Methods}\label{sec:desp.method}
In this section, we give a short description of candidate methods that will be used for the construction of the variability map. We first consider a multiple linear regression model, which is a simple model. We also consider several popular classes of surrogate models as the candidate methods, which are the Gamma generalized linear model (GLM) with a polynomial structure, the linear Shepard (LSP) method that is based on inverse distance weight, the multivariate adaptive regression splines (MARS), Gaussian process based methods, tree based models such as the Bayesian additive regression trees (BART), and categorical Gaussian process (CGP). These methods are flexible and can automatically capture globally complex relationships, potentially uncovering interactions between variables that have dominant effects on IO variability.

\textbf{Linear Model.} For the linear model (LM), the response $y_i$ is modeled as
$$y_i=\beta_0+\xvec_i'\betavec+\varepsilon_i,$$
where the variability map $f(\xvec)$ is modeled as $f(\xvec)=\beta_0+\xvec'\betavec$, $\beta_0$ and $\betavec$ are the regression coefficients, and $\varepsilon_i$ is the error term. The parameter estimation is done by using the least-squares method.

\textbf{Gamma GLM.} In some applications, the LM can be insufficient for modeling standard deviation as mentioned in Chapter~11 of \citeN{MyersMontgomeryAnderson-Cook2016}. Hence, we further consider a Gamma GLM (GAMGLM) with a polynomial mean structure. In the GAMGLM setting, we assume $y_i$ follows a Gamma distribution with mean $\mu_i$. Through the log link function, the mean $\mu_i$ is linked to the linear predictor $\eta_i$ as $\mu_i=\log(\eta_i)$. We use a polynomial structure for $\eta_i$ as follows for the HPC dataset,
$$\eta_i = \beta_0 + \sum_{j=1}^4 \beta_{j} x_{ij} +\sum_{j=3}^4 \gamma_{j} x_{ij}^2 + \sum_{j = 3}^4\sum_{k = 1}^{j-1}\beta_{jk} x_{ij} x_{ik}+\sum_{j=3}^4 \delta_{j} x_{ij}^3.
$$
Here, the covariates are defined in \eqref{eqn:x.def} and $\{\beta_0, \beta_j, \gamma_j, \beta_{jk}, \delta_j\}$ are regression coefficients. Note that we only consider the linear terms for $x_{i1}$ (Log of File Size) and $x_{i2}$ (Log of Record Size) because there are only six unique combinations for the two variables. The variability map $f(\xvec)$ is estimated as $\mu$, which is the mean of the Gamma distribution at point $\xvec$. The parameters can be estimated by using the maximum likelihood method.

\textbf{The Linear Shepard Method.} The LSP method is an approximation algorithm based modification of the original Shepard algorithm (e.g., \shortciteNP{Thacker}). The LSP approximation is constructed using linear combinations of locally linear fits. To introduce LSP, we need to define some notation first. Let $p$ be the length of $\xvec$, $n$ be the number of data points and $m=\min\{n, 3p/2\}$ as set in \shortciteN{Thacker}. Let $d_i(\xvec)=\|\xvec-\xvec_i\|_2$ be the distance between $\xvec$ and $\xvec_i$ and let $d=\max_{i,j}d_i(\xvec_j)$ be the maximum distance among all pairs of $\xvec_i$. Let $r_i$ be the smallest radius of the closed ball around $\xvec_i$ such that at least $m$ number of data points are included in the ball. Let $u_i=\min\{d/2, r_i\}$ and $v_i=1.1r_i$, also as designed in \shortciteN{Thacker}.

In LSP, for an arbitrary point $\xvec$, the variability $f(\xvec)$ is estimated as $\wh{f}(\xvec)$. In particular,
$$
\wh{f}(\xvec)= \frac{\sum_{i=1}^{n} w_i(\xvec)p_i(\xvec)}{\sum_{i=1}^{n} w_i(\xvec)},
$$
where $w_i(\xvec)=\{[1/d_i(\xvec)-1/u_i]_+\}^2$. That is, the weight is set as $w_i(\xvec)$ for those $\xvec$'s are within the ball for $\xvec_i$ with radius $u_i$ and is set to 0 for those $\xvec$'s are outside the ball for $\xvec_i$. Here $[x]_+=x$ if $x\geq0$ and $[x]_+=0$ if $x<0$.

We use the locally weighted least-squares fit to obtain the $p_i(\xvec)$ with the following form,
$p_i(\xvec)=y_i+(\xvec-\xvec_i)'\betavec_i,$ where $\betavec_i$ is the vector for coefficients. The weight used in the locally weighted least-squares fit is defined as $\omega_{ij}=\{[1/d_i(\xvec_j)-1/v_j]_{+}\}^2$. That is, the weight is set as $\omega_{ij}$ if $\xvec_i$ is within the ball for $\xvec_j$ with radius $v_j$ and is set to zero if $\xvec_i$ is outside the ball for $\xvec_j$. Then, the coefficient $\betavec_i$ can be obtained by solving the weighted least-squares problem.

\textbf{Multivariate Adaptive Regression Splines (MARS).} Let $\xvec_i=(x_{i1},\cdots, x_{ij},\cdots, x_{ip})'$. For any $\xvec=(x_1,\cdots, x_j, \cdots, x_p)'$, MARS bases are defined as, $\mathcal{C}=\left\{\left(x_j-t\right)_+, \left(t-x_j\right)_+\right\}$, for $t \in \; \{x_{1j},\; x_{2j}, \dots,\; x_{nj}\},\; j=1,\; 2,\; \dots,\; p.$ Using the MARS method, the variability map $f(\xvec)$ is estimated as,
\begin{align*}
\wh{f}(\xvec)=\wh\beta_0+\sum_{l=1}^m \wh\beta_lh_l(\xvec),
\end{align*}
where $\beta_0$ and $\beta_l$ are coefficients, and $h_l(X)$ is a function in $\mathcal{C}$ or a product of two or more functions in $\mathcal{C}$. A step-wise procedure is used for the selection of bases from the set $\mathcal{C}$. One starts with constant function $h_0(\xvec)=1$ in the model. At each step, one adds a term of the following form which minimizes the squared errors most to the current model,
\begin{gather*}
\wh{\beta}_{m+1}h_l(\xvec)\cdot(x_j-t)_++\wh{\beta}_{m+2}h_l(\xvec)\cdot(t-x_j)_+,
\end{gather*}
where $h_l(\xvec)$ are the terms in the current model. At the end of the step-wise procedure, a large model may result and thus a backward procedure is performed to eliminate some terms. To determine the best model, the generalized cross-validation procedure is used.

\textbf{Bayesian Treed Gaussian Process.} Treed Gaussian processes (TGP) are widely used methods in modeling and prediction of data collected from computer experiments. In general, TGP uses Bayesian methods to build a CART model, and it partitions the parameter space into disjoint regions using a tree structure. In each region, a local Gaussian process is fit to the data from that region. Thus, a TGP typically consists of a tree structure construction and local Gaussian process fitting, which provide the potential to capture complex relations between system configuration $\xvec$ and the performance variability $f(\xvec)$. We consider three types of TGP in this paper, which are TGP with a linear trend (denoted as TGPlm), TGP with a constant mean (denoted as TGPcart), and dynamic trees (denoted as DynaTree).

The treed partition models use binary splits on the value of univariate variables to divide the input space. Each partition is recursive, and a new partition can occur in previously divided regions. Other tree construction methods such as swapping, pruning, rotation can take place after the splitting.

For a specific region, the data from a specific region $j$ are denoted by $\Xvec_{j}$ and $\yvec_{j}$, where each row of $\Xvec_{j}$ contains one system configuration and $\yvec_{j}$ is a vector for the corresponding outputs from region $j$. For each region, we use the following Gaussian process model,
\begin{align*}
\yvec_{j}\sim \NOR[\mu(\Xvec_{j}, \thetavec_{j}), \Sigmavec_{j}].
\end{align*}
Here $\mu(\Xvec_{j}, \thetavec_{j})$ is the mean structure which is a function of $\Xvec_{j}$ with unknown parameters $\thetavec_{j}$, and $\Sigmavec_{j}$ is the variance-covariance matrix. With the specification of priors, the parameters can be estimated through Gibbs sampling. The estimation and prediction of $f(\xvec)$ at any $\xvec$ can be done using the posterior distribution.

TGPlm uses a linear trend Gaussian process in local partition while TGPcart uses a constant mean. Dynamic trees can explore posterior tree space by stochastically proposing incremental modifications to the tree structure (such as splitting, pruning, and swapping) in each terminal leaf.

\textbf{Bayesian Additive Regression Trees.} The BART consists of two components, which are a sum-of-trees model and a regularization prior on the sum-of-tree model parameters. For a BART model with $m$ trees, the variability map $f(\xvec)$ can be modeled as
$$y_i=f(\xvec_i)+\varepsilon_i=\sum_{j=1}^{m} g(\xvec_i;\,T_j;\,\thetavec_j)+\varepsilon_i.$$
Here the error term $\varepsilon_i\sim \NOR(0,\sigma^2)$, and $T_j$ is a binary regression tree that partitions the parameter space. The associated terminal leaf parameters for tree $T_j$ is denoted by $\thetavec_j$. The function $g(\xvec;\,T_j;\,\thetavec_j)$ provides the output values for $\xvec$ according to tree $T_j$. With the specification of the prior distributions, the Bayesian back-fitting MCMC algorithm can be used for model estimation and prediction.

\textbf{Gaussian process with qualitative and quantitative factors.} The categorical Gaussian process (CGP) is used to handle qualitative and quantitative factors input variables, which is a modification to the classic Gaussian process. We use the CGP model in \citeN{ZhouQianZhou2011}. Starting from the original Gaussian process model, $\yvec \sim \textrm{N}\left[\mu,\Sigmavec(\Xvec,\thetavec)\right]$,
where $\mu$ is the global mean, $\Xvec$ is the design matrix for continuous variables, and $\Sigmavec(\Xvec,\thetavec)$ is the covariance matrix with parameters $\thetavec$. We consider two categorical variables, the IO scheduler and the VMIO scheduler, as the categorical input variables, in addition to those continuous variables. With the consideration of the IO scheduler and the VMIO scheduler, the length of $\yvec$ is $805\times3\times3\times3=7245$. Due to computational constraints, we still model the levels of the IO Operation Mode separately. Thus, we have 13 separate variability maps under CGP model.

Let $\zvec^{io}$ and $\zvec^{vmio}$ be the corresponding coded vectors for IO scheduler and VMIO scheduler (i.e., CFQ, 1; DEAD, 2; and NOOP, 3). Let $z^{io}_i$ and $z^{vmio}_i$ be the corresponding $i$th elements in $\zvec^{io}$ and $\zvec^{vmio}$, which takes values in $\{1,2,3\}$. For CGP model, the covariance between two points $(\xvec_i', z^{io}_i, z^{vmio}_i)'$ and $(\xvec_j', z^{io}_j, z^{vmio}_j)'$ is further modeled as,
$$\Sigmavec(\Xvec,\thetavec)_{ij}\times\tau^{io}_{z^{io}_iz^{io}_j}
\times\tau^{vmio}_{z^{vmio}_iz^{vmio}_j},
$$
where $\tau^{vmio}_{kl}$ and $\tau^{io}_{kl}, k=1,2,3, l=1,2,3$ are correlation parameters to be estimated. All the parameters are estimated by using the maximum likelihood method.

\subsection{Estimation and Visualization of Variability Maps}
In this section, we use the training set in Table~\ref{tab:exp.setup} to estimate the variability map. The candidate methods described in Section~\ref{sec:desp.method} are used. The log-transform is applied to the response variable.  For the linear regression model, we use four predictors, namely, the log of File Size, the log of Record Size, the log of the Number of Threads, and the CPU frequency. For MARS, the order of the highest basis function is set to three and the exclusive method is used to add new basis functions in each iteration. For the DynaTree method, the linear terminal Gaussian process is used.

\begin{figure}
\begin{center}
\begin{tabular}{cc}
\includegraphics[width=.45\textwidth]{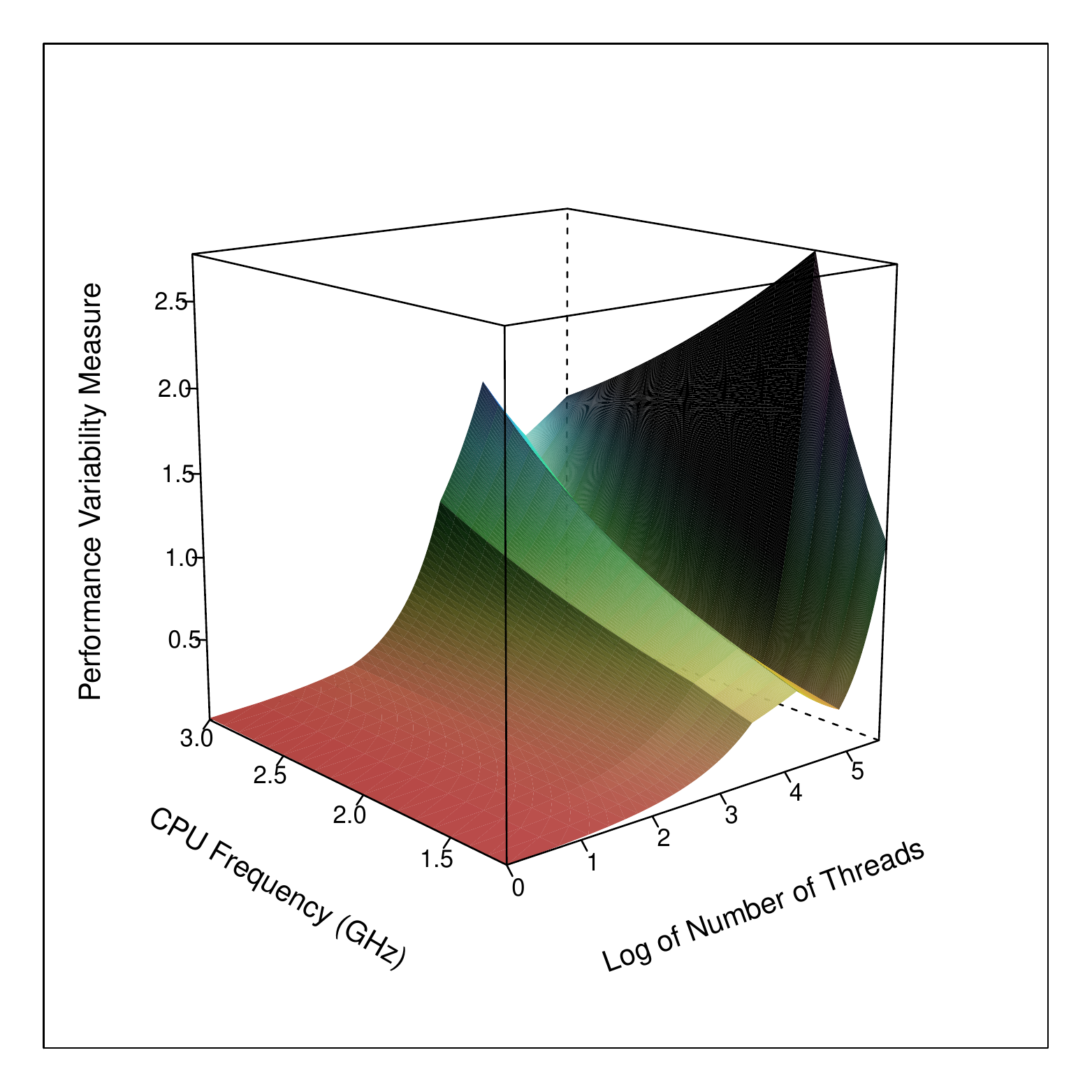}&
\includegraphics[width=.45\textwidth, angle=90]{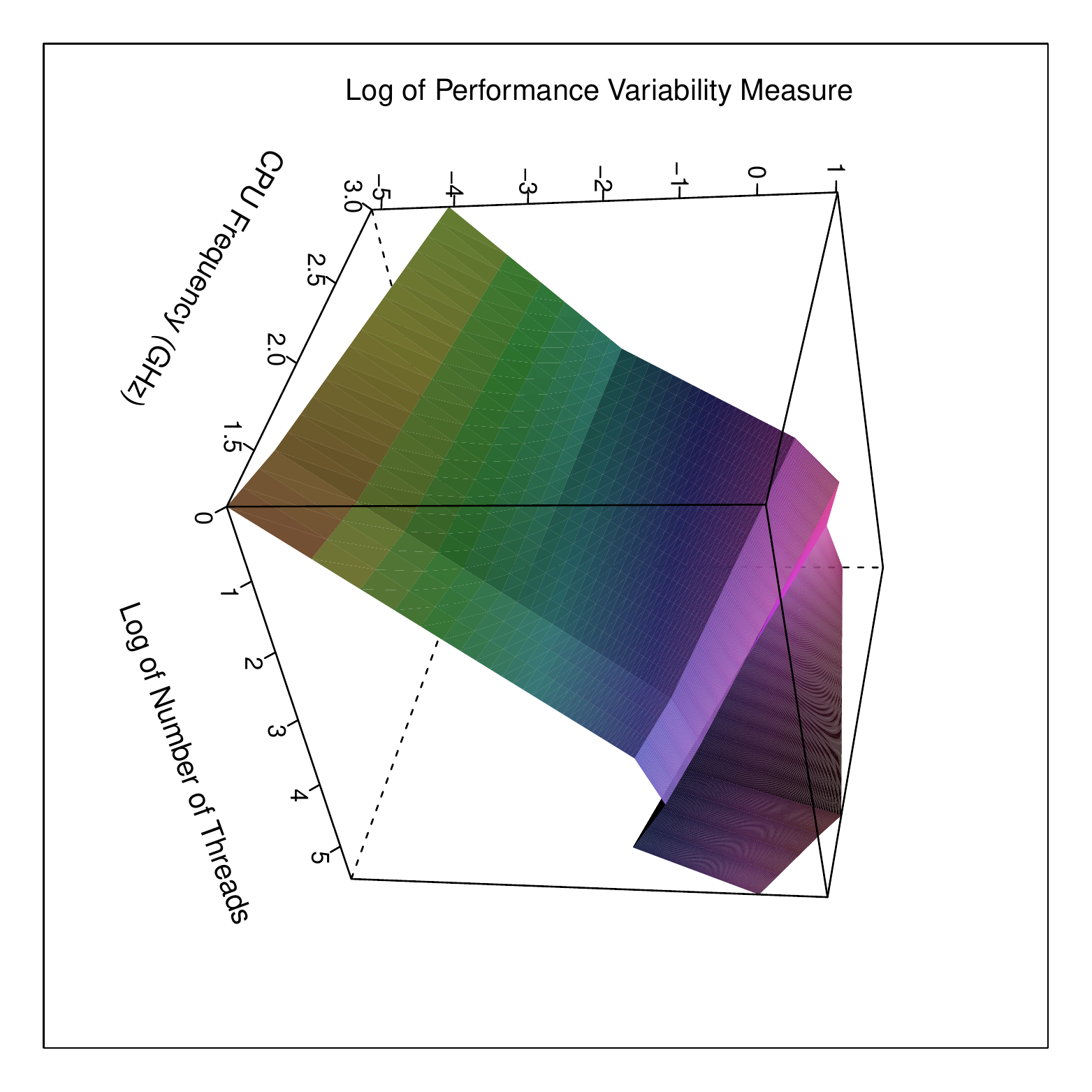}\\
(a) PVM& (b) Log of PVM
\end{tabular}
\end{center}
\caption{Perspective plots of estimated variability map as a function of CPU Frequency and the Number of Threads, using the MARS algorithm. The IO Operation Mode is Fwrite, IO Scheduler is CFQ, the VM IO Scheduler is NOOP, the File Size is 1024, and the Record Size is 512. }\label{fig:est.vmap}
\end{figure}

For each distinct combination of these categorical variables, we generate separate variability maps, resulting in 117 variability maps, except for CGP which has 13 variability maps. Each variability map is a four-dimensional surface. To visualize the estimated variability map, Figure~\ref{fig:est.vmap} shows two perspective plots of estimated variability maps as a function of CPU Frequency and the Number of Threads, using the MARS algorithm. The system configuration of Figure~\ref{fig:est.vmap} is the same as in Figure~\ref{fig:pesp}, with IO Operation Mode fixed at Fwrite, IO Scheduler at CFQ, VM IO Scheduler at NOOP, File Size at 1024, and Record Size at 512. From the figure, we see that the estimated variability map by MARS can capture most patterns in raw data as shown in Figure~\ref{fig:pesp}.

\section{Prediction and Method Comparisons}\label{sec:pred.meth.comp}
\subsection{Criteria for Comparisons}
In this section, we conduct extensive comparisons to study the properties of existing methods. In particular, we are interested in whether the models can predict the performance variability well for a new system configuration. To evaluate the performance of the candidate methods, we need to discuss our comparison criteria first. Let $y_k$ be the true value and $\yhat_k$ be the predicted value, for $k=1,\cdots, n$. Here $n$ is both 585 for the interpolation and extrapolation test sets, after combining the points from all variability maps. The root mean squared error (RMSE) is defined as $\sqrt{\sum_{k=1}^{n}(\yhat_k-y_k)^2/n}$, while the mean absolute error (MAE) is defined as $\sum_{k=1}^{n}|\yhat_k-y_k|/n$. The RMSE and MAE have the same scale as the original data. Several other relative errors are defined as,
$$\ER_1=\frac{\sum_{k=1}^{n}|\yhat_k/y_k-1|}{n}, \quad \ER_2=\frac{\sqrt{\sum_{k=1}^{n}(\yhat_k-y_k)^2/n}}{\sum_{k=1}^{n}y_k/n}, \quad \textrm{ and }\quad \ER_3=\frac{\sum_{k=1}^{n}|\yhat_k-y_k|/n}{\sum_{k=1}^{n}y_k/n}.$$

Note that the $\ER_1$ is the average of the point-wise error rate. However, $\ER_1$ tends to be large when the true value is small. The $\ER_2$ is the RMSE divided by the mean $\sum_{k=1}^{n}y_k/n$ to provide a scale-free measure of error. However, RMSE tends to be dominated by those terms with large errors. Note that $\ER_2$ is also referred to as the coefficient of variation in the statistical literature. The $\ER_3$ is the MAE divided by the mean to provide a scale-free measure. Because each error measure has its pros and cons, we will present our results under different error measures to provide a comprehensive comparison.

\subsection{Interpolation}
In this section, we make a comparison for predictability of different methods when a new design point lies within the input space (i.e., interpolation). Using the training set with 94185 data points, variability maps are constructed by using the candidate methods. The log transformation for the thread count, file size, record size and responses is used. For each variability map, there are five points to be predicted, as shown in Table~\ref{tab:test.set.setup}. In total there are $117\times5=585$ points in the test set for interpolation.

\begin{figure}
\begin{center}
\includegraphics[width=.55\textwidth]{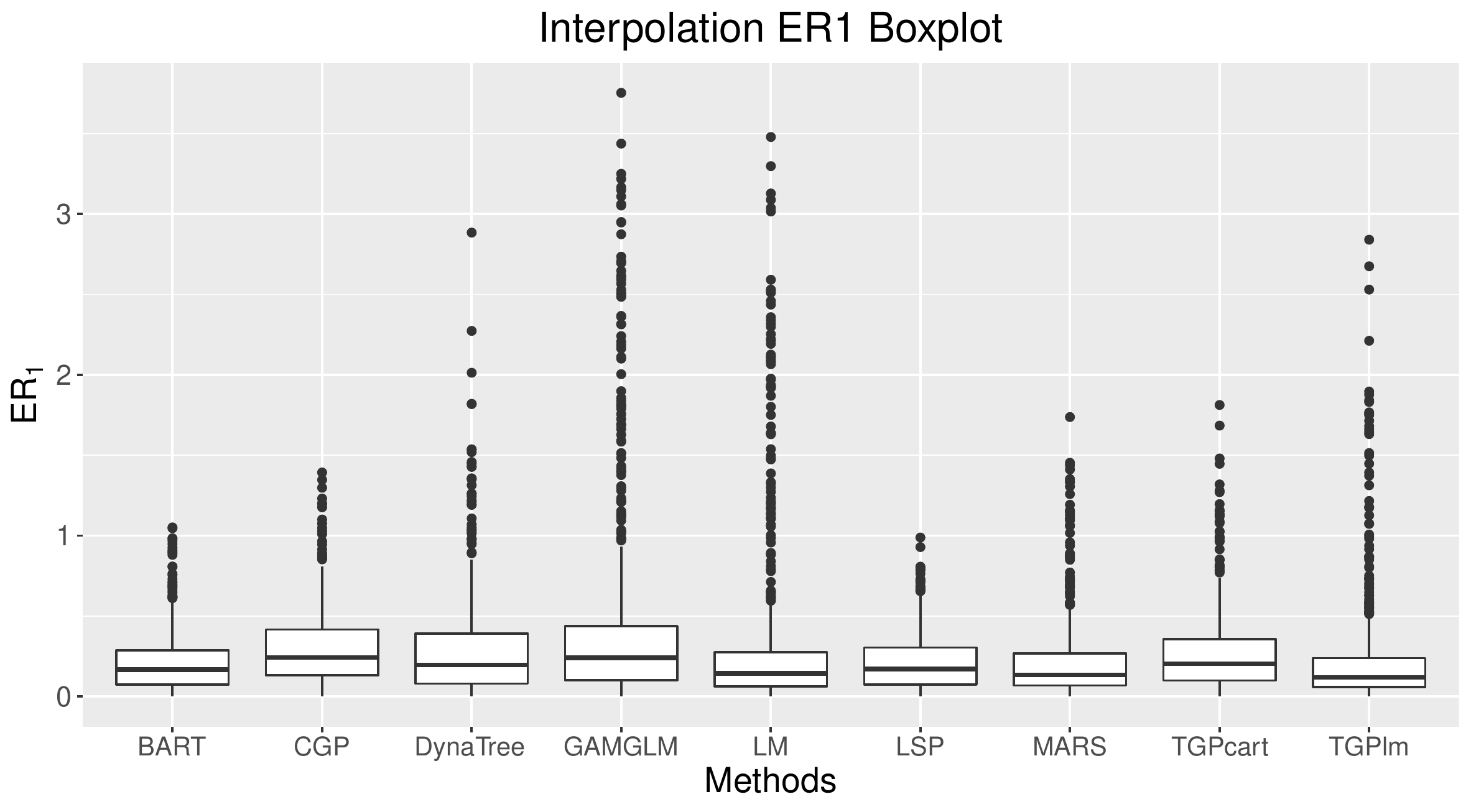}
\end{center}
\caption{Box plot for the 585 individual relative error rates for the candidate methods based on the interpolation test set. }\label{fig:int.err.box.plot}
\end{figure}

Various error measures are then computed. Table~\ref{tab:error.rate.int} shows the RMSE, MAE, and three error rates for the candidate methods based on the interpolation test set. The RMSE and MAE are in the magnitude of $10^{7}$. Even under different error measures, the MARS method turns out to be the best one resulting in the smallest error rate. The LSP and BART generate results that are best after MARS. The three TGP based methods and CGP are in the middle based on $\ER_1$. The LM and GAMGLM tend to have large errors based on $\ER_1$ but their error rates are smaller based on $\ER_2$ and $\ER_3$. To investigate the spread of error rates, Figure~\ref{fig:int.err.box.plot} shows a box plot for the 585 individual relative error rates ($\yhat_k/y_k$) for the candidate methods. One can see that the LSP method has the least spread, and MARS and TGPlm have the least median error rates. Overall, the LSP method tends to work well for interpolation prediction problems as it has the least error spread and a relatively low error.

\begin{table}
\begin{center}
\caption{The RMSE, MAE, and three error rates for the candidate methods based on the interpolation test set. The RMSE and MAE are in the magnitude of $10^{7}$.}\label{tab:error.rate.int}
\vspace{.5em}
\begin{tabular}{l|c|cc|cc}\hline\hline
Method	  & $\ER_1$   & RMSE  & $\ER_2$   & MAE	  & $\ER_3$  \\\hline
LM	      & 0.357 & 1.135 & 0.227 & 0.827 & 0.165\\
GAMGLM    & 0.481 & 1.407 & 0.281 & 1.116 & 0.223\\
LSP	      & 0.216 &	1.269 & 0.254 & 0.891 & 0.178\\
MARS	  & 0.223 &	0.913 & 0.183 & 0.695 & 0.139\\
TGPlm	  & 0.245 & 1.024 & 0.205 & 0.707 & 0.141\\
TGPcart	  & 0.268 &	1.519 & 0.304 & 1.072 & 0.214\\
DynaTree  & 0.296 &	2.462 & 0.492 & 1.305 & 0.261\\
BART	  & 0.220 &	1.222 & 0.244 & 0.878 & 0.175\\
CGP	  	  & 0.298 &	1.770 & 0.354 & 1.300 & 0.261\\
\hline\hline
\end{tabular}
\end{center}
\end{table}

\subsection{Extrapolation}
In this section, we make a similar comparison for predictability of different methods when a new design point falls outside the convex hull of the input space (i.e., extrapolation). Similar to the setting of interpolation tests, for each variability map, there are five points to be predicted, as shown in Table~\ref{tab:test.set.setup}. In total there are $117\times5=585$ points in the test set for extrapolation. We find that the models with log transformed responses tend to generate nonsensically large errors for extrapolation due to the effect of exponentiation. Thus, for the extrapolation, we model the response at its original scale.

\begin{figure}
\begin{center}
\includegraphics[width=.55\textwidth]{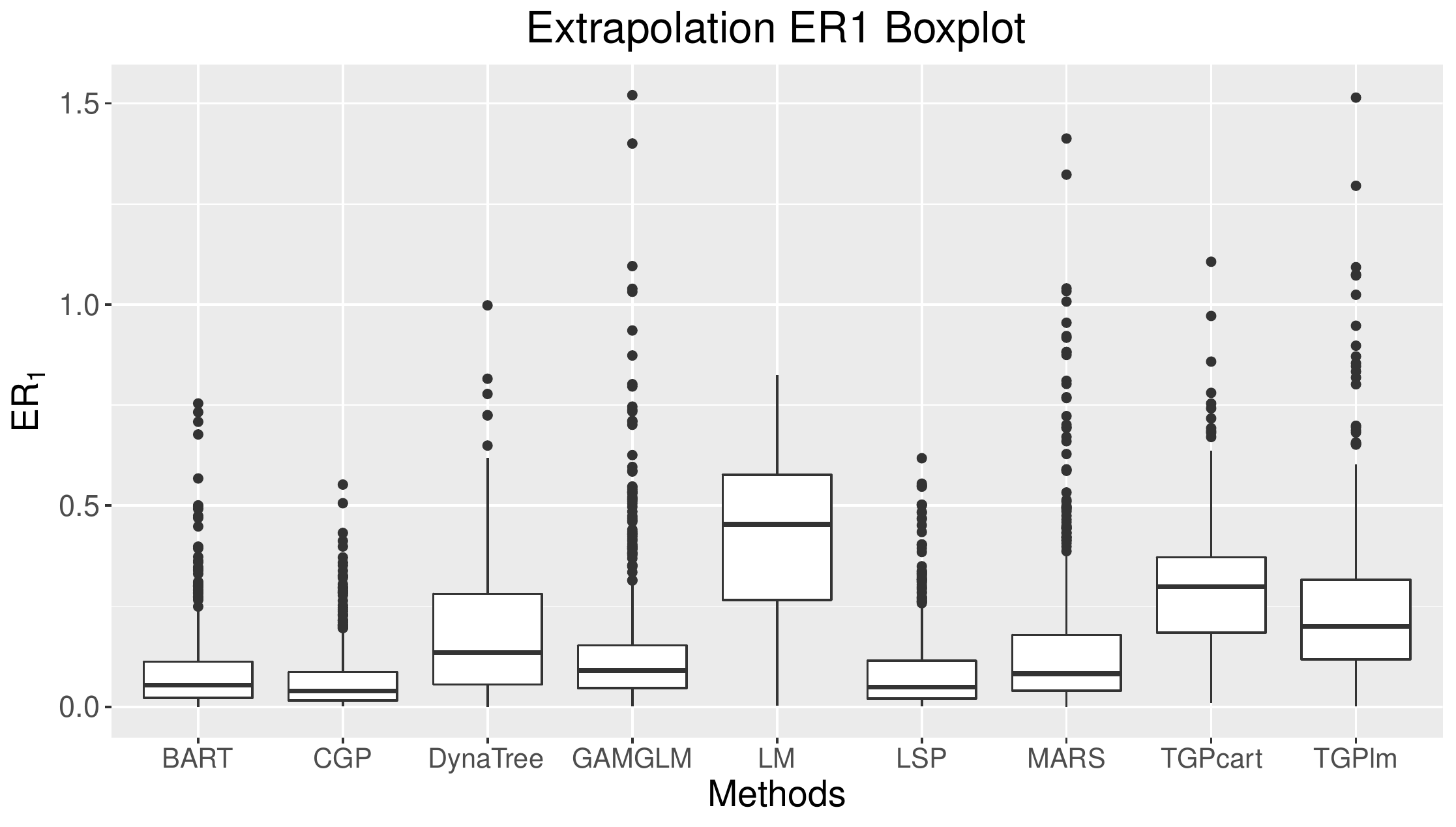}
\end{center}
\caption{Box plot for the 585 individual relative error rates for the candidate methods based on the extrapolation test set. }\label{fig:extra.err.box.plot}
\end{figure}

Table~\ref{tab:error.rate.ext} shows the various error rates based on the extrapolation test set. Among all error measures, the CGP shows the best result. The BART and LSP methods are the next after CGP. The MARS and GAMGLM methods are in the middle, and the TGP methods tend to generate large errors. Figure~\ref{fig:extra.err.box.plot} shows the box plot for the 585 individual relative error rates. The figure shows that the CGP and LSP methods have the least amount of spread in terms of individual errors. Most surrogate models in computer experiments are built for interpolation and their extrapolation ability is not well studied. Surprisingly, we find that both the CGP and LSP methods have good extrapolation ability under our current test setting. Overall, the CGP shows potential as a good extrapolation technique. However, one should note that extrapolation is always a difficult task and should be exercised with caution.

\begin{table}
\begin{center}
\caption{The RMSE, MAE, and three error rates for the candidate methods based on the extrapolation test set. The RMSE and MAE are in the magnitude of $10^{7}$.}\label{tab:error.rate.ext}
\vspace{.5em}
\begin{tabular}{l|c|cc|cc}\hline\hline
Method	  & $\ER_1$   & RMSE    & $\ER_2$   & MAE	    & $\ER_3$  \\\hline
LM	      & 0.412 & 7.509   & 0.712	& 5.272     & 0.500\\
GAMGLM	  & 0.142 &	1.490   & 0.141 & 1.087     & 0.103\\
LSP	      & 0.089 & 1.402   & 0.094	& 0.634     & 0.060\\
MARS	  & 0.148 & 0.995   & 0.133	& 0.952     & 0.090\\
TGPlm     & 0.236 & 3.851   & 0.365 & 2.561     & 0.243\\
TGPcart   & 0.289 & 4.701   & 0.446 & 3.376     & 0.320\\
DynaTree  & 0.184 & 3.390   & 0.321 & 1.927     & 0.183\\
BART      & 0.087 & 0.946   & 0.090 & 0.621     & 0.059\\
CGP	  	  & 0.066 &	0.630 	& 0.060 & 0.434 	& 0.041\\
\hline\hline
\end{tabular}
\end{center}
\end{table}

\section{Management of Performance Variability}\label{sec:optim.per.var}
\subsection{The System Optimization Problem}
In this section, we address the problem of variability management. One method for balancing overall performance with performance variability is to minimize performance variability, while maintaining acceptable overall performance. Let $\xvec$ be the system configuration, and $\wh{f}(\xvec)$ be the estimated variability map. To do the system design, one also needs to know the performance surface $m(\xvec)$ at any $\xvec$. The performance surface is the mean throughput at $\xvec$. With the training set, $m(\xvec)$ can be estimated by using the methods discussed in Section~\ref{sec:desp.method}, which is denoted by $\wh{m}(\xvec)$. Let $\D$ be the feasible domain of $\xvec$. Let $m_0$ be the minimum performance requirement. The system optimization problem can be formulated as the following optimization problem. That is
\begin{align}\label{eqn:min.prob}
\min_{\xvec\in\D} \wh{f}(\xvec), \quad {\textrm{subject to}} \quad \wh{m}(\xvec)\geq m_0.
\end{align}
Based on the results in the comparison study, we build the system variability map using LSP, and the performance surface is estimated using LSP as well, due to its good prediction performance and computational efficiency. For illustration, the performance requirement threshold $m_0$ is specified as the mean of the throughput from the 805 points in the training set for each variability map. One can, however, specify any values for the threshold.

\begin{figure}
	\centering
	\includegraphics[width=1\textwidth, angle=0]{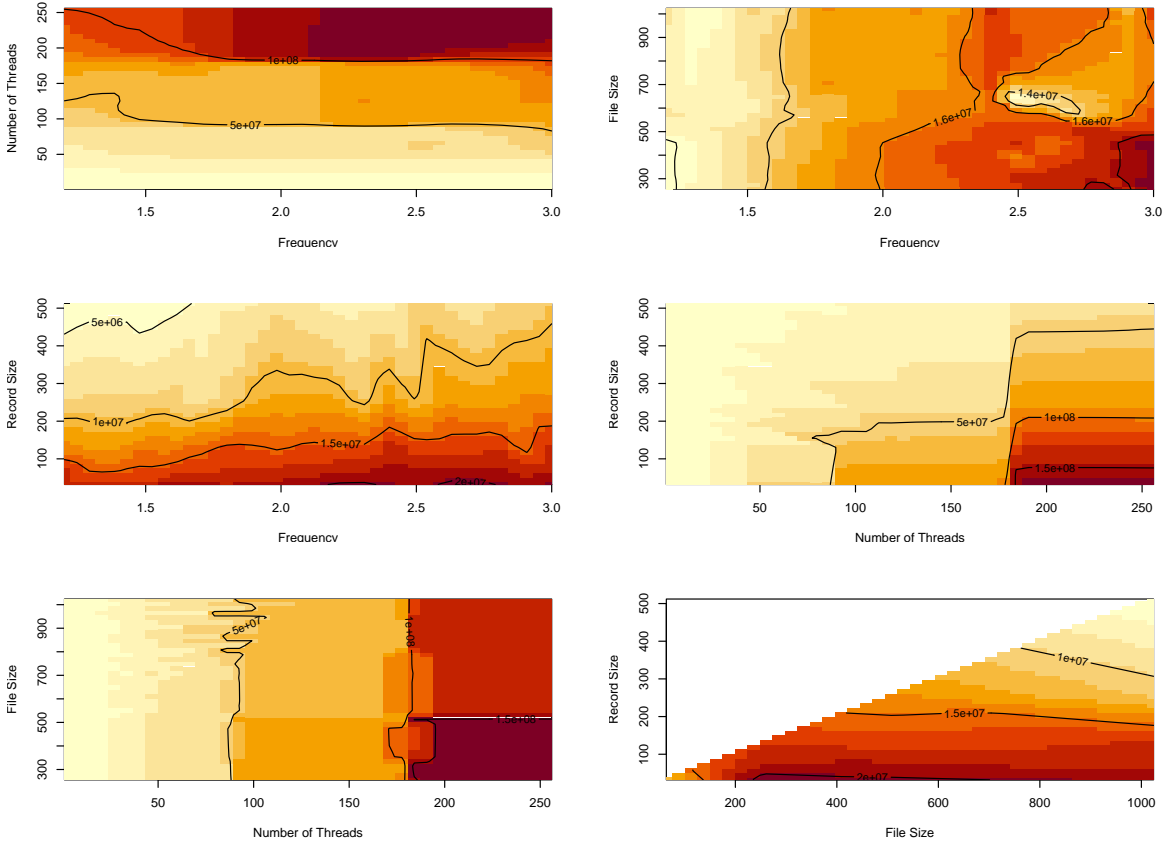}	
\caption{Contour plots of $\wh{f}(\xvec)$ under the configuration that File Size is 1024, Record Size is 128, Frequency is 2.4, Number of Threads is 32, Operation Mode is Fread, IO Scheduler is DEAD and VM IO Scheduler is CFQ.}\label{fig:3dsurface}
\end{figure}

The objective function $\wh{f}(\xvec)$ is a 4-dimensional surface. To visualize the objective function, we show the 2-dimensional cross sections of the 4-dimensional surface. That is, one fixes two of the four dimensions (e.g., File Size and Record Size), and draws the contour of $\wh{f}(\xvec)$ as a function of the rest two dimensions (e.g., the Number of Threads and CPU Frequency). Figure~\ref{fig:3dsurface} shows the contour plots of the cross sections of the objective function under the configuration: File Size of 1024, Record Size of 128, Frequency of 2.4, Number of Threads of 32, Operation Mode of Fread, IO Scheduler of DEAD and VM IO Scheduler of CFQ. The plots in Figure~\ref{fig:3dsurface} show that the objective function is complicated.

\subsection{Methods for System Configuration Optimization}
The optimization problem is a challenging problem due to the complexity of the multi-dimension surface. In this paper, we use the augmented Lagrange method (e.g., \citeNP{Hestenes1969}) to solve the optimization problem in \eqref{eqn:min.prob}. By introducing a slack variable $z$, \eqref{eqn:min.prob} is equivalent to
\begin{align*}
\min_{\xvec\in\D, \,\,z} \wh{f}(\xvec), \quad {\textrm{subject to}} \quad m_0-\wh{m}(\xvec)+z^2= 0.
\end{align*}
Using the augmented Lagrange method, we have the following new objective function,
\begin{align}\label{eqn:aug.lag}
L(\xvec,z,u,c) = \wh{f}(\xvec)+u\left[m_0-\wh{m}(\xvec)+z^2\right]+\frac{c}{2}\left[m_0-\wh{m}(\xvec)+z^2\right]^2,
\end{align}
where $u$ is the Lagrange multiplier and $c$ is an additional parameter introduced by the algorithm. Note that $\xvec$ and $z$ are separable. We first minimize $z$ and obtain,
\begin{align*}
L_z(\xvec,u,c) = \wh{f}(\xvec)+\frac{1}{2c}\max\{0,u+c[m_0-\wh{m}(\xvec)]\}^2-u^2.
\end{align*}
In theory, the problem in \eqref{eqn:min.prob} can be solved in an iterative way. One can pick a sequence of values for $c$, denoted by $c_k$, such that $\lim_{k\rightarrow \infty}c_k = \infty$. Then $\xvec_k$ and $u_k$ can be updated in the following way, $\xvec_k = \arg\min L_z(\xvec,u_k,c_k)$ and
$u_{k+1} = \max\{0, u_k+c_k\left[m_0-\wh{m}(\xvec_k)\right]\}$. The sequence $\xvec_k$ will approach to the solution for \eqref{eqn:min.prob}. The iteration stops when the distance between two consecutive updates of $\xvec_k$ is small enough (e.g., $1e-6$).

In this paper, we use the VTDIRECT95 package (\citeNP{vtdirect95}) which can solve the problem with one global iteration run.  We set the $c$ parameter to be a large value (e.g., $c=1000$), and use VTDIRECT95 to solve~\eqref{eqn:aug.lag} directly.

Let $\xvec^{\ast}$ be the solution to \eqref{eqn:min.prob}. Due to some system requirements, some adjustments are needed for $\xvec^{\ast}$. We need to find proper file size and record size such that the ratio of file size and record size is an integer. For a given pair of file size and record size, we find the nearest corrected file size and record size such that the corrected file size and record size are integers, and the ratio of corrected file size and record size is also an integer. We also round the number of threads to the nearest integer.

\subsection{System Optimization Results}
For system design optimization, we aim to find the best system configurations (i.e., IO Scheduler, VM IO Scheduler, File Size, Record Size, Number of Threads, and CPU Frequency) under each operation mode. For each operation mode, there are nine IO and VM IO scheduler combinations, we optimize the nine configurations separately and select the IO and VM IO schedulers with the least cost function value.

Table~\ref{tab:optim} shows the optimal system configurations for the 13 operation modes. We found that most write operations need a high number of threads to have optimal performance in terms of stability and speed. While for the read operations, the system has the best performance when the number of threads is 16, which is the number of physically available threads from the server. When the number of threads is more than the number of the physically available threads, the system typically shows more variability. The optimal file size doubles the record size, which means smaller file size and record size ratio leads to smaller variability. Lower frequencies typically lead to smaller variability, given that the requirement for the performance is met. The results in Table~\ref{tab:optim} can provide useful information for configuring new systems.

\begin{table}
	\centering
	\caption{Examples of system configurations for each operation mode after optimization to minimize performance variability subject to minimum performance requirements. Here ``Fs'' means File Size, ``Rs'' means Record Size, and ``Freq'' means the CPU Frequency.}\label{tab:optim}
\vspace{1.5ex}
	\begin{tabular}{c|c|c|c|c|c|c}
		\hline\hline
		Mode            & IO   & VMIO   & Fs     & Rs     & Thread & Freq     \\
		\hline
		        Fread&  CFQ&  CFQ &         752 &         376 &     16& 1.73   \\\hline
		       Fwrite& DEAD&  CFQ &          64 &          32 &     30& 1.33   \\\hline
		 Initialwrite& DEAD& NOOP &          64 &          32 &     30& 1.31   \\\hline
		Mixedworkload&  CFQ& NOOP &         780 &         390 &     16& 1.97   \\\hline
		        Pread& DEAD&  CFQ &         704 &         352 &     16& 1.20   \\\hline
		       Pwrite& DEAD&  CFQ &         712 &         356 &     30& 1.20   \\\hline
		   Randomread& NOOP& NOOP &        1024 &         512 &     16& 1.34   \\\hline
		  Randomwrite& NOOP& NOOP &         308 &         154 &     16& 1.73   \\\hline
		      Re-read& NOOP& DEAD &         848 &         424 &     16& 1.70   \\\hline
		         Read& NOOP& NOOP &         752 &         376 &     16& 1.20   \\\hline
		  ReverseRead&  CFQ& NOOP &         990 &         495 &     16& 1.30   \\\hline
		      Rewrite& DEAD& NOOP &         832 &         416 &     16& 1.88   \\\hline
		   Strideread& DEAD& DEAD &         752 &         376 &     16& 1.66   \\\hline\hline
	\end{tabular}
\end{table}

\section{Concluding Remarks and Areas for Future Research}\label{sec:conclusion}
In this paper, we investigate HPC variability using various statistical techniques.  We fit the variability data with different surrogate models. We show that the MARS, BART, and LSP methods have good performance in interpolation, and the CGP and LSP methods have good performance in extrapolation, under several error measures. The users can choose the corresponding methods based on the purpose of the analysis (i.e., interpolation versus extrapolation). Overall, the LSP method has small $\ER_1$ error rates with small error spreads for both scenarios. In addition, we also develop an optimization procedure to manage system variability and select the best system configuration under different modes. The tools developed in this paper can be useful for the practice of HPC variability management and HPC system building.

We want to point out that our paper serves as an initial step in analyzing HPC performance variability statistically. We demonstrate the usefulness of the particular prediction methodologies in understanding the variability in IO. However, we want to emphasize that more extensive work is needed in the future for studying HPC performance variability. Here we discuss some possible areas for future research. We currently use a grid-based design for selecting design points in the data collection stage. In the future, we can consider sequential design techniques to select the design points. Space filling design techniques can also be considered for data collection (e.g., \citeNP{Joseph2016}).

Even though we can predict performance variability well, it is still important to understand the root causes of  variability. In the future, it will be interesting to collect more data, such as the performance counter statistics which provide finer details on the IO operation, and analyze such data to reveal the root causes of performance variability. Variable selection techniques will be useful in finding important factors in performance variability.

We focus on the prediction of the standard deviation of the throughput in this paper. \shortciteN{Xuetal-JPDC-2020} show that the distribution of the throughput is multi-modal and thus it is complicated. A more ambitious goal is to predict the system throughput distribution generally (e.g., \shortciteNP{lux2018nonparametric}). \citeN{HungJosephMelkote2015} developed a Gibbs sampling-based expectation-maximization algorithm for efficiently fitting a Kriging model to the functional responses. Techniques in functional data analysis can also be considered in modeling functional responses. In literature, single index models are used in Gaussian processes to make predictions in the context of computer experiments (e.g., \citeNP{ChoiShiWang2011}, and \citeNP{GramacyLian2012}). It will be interesting to consider Gaussian processes with single index models to predict performance variability in HPC setting.

In this paper, we consider a Gaussian process with categorical and continuous inputs, which can handle two categorical variables (i.e., the IO scheduler and the VMIO scheduler). However, we found estimating a model using all categorical variables (i.e., including IO Operation Mode) to be challenging. Because the IO Operation Mode has 13 levels, in order to show all interactions, we need to estimate $13\times12/2=78$ parameters. In addition, all observations will be correlated (i.e., we need to invert a covariance matrix with dimension $94185\times94185$). In the future, it will be interesting to investigate methods that can handle this large scale of data.

The developed data analytical framework for variability can be applied to areas other than HPC in computer science. The two major areas are cloud computing and system security. In cloud computing, large scale clouds operated by large corporations seek to provide a guaranteed level of service to a client. The variability management framework presented in this paper provides new opportunities for tradeoffs between system stability and performance. In system security, malware attacks may impact system performance enough to create noticeable changes in performance variability. Variability management can potentially increase the effectiveness of these types of malware detection techniques.

\section*{Supplementary Materials}
The following supplementary materials are available online.
\begin{description}
\item[Code and data:] Computing code for data analysis, comparisons, and optimization. The HPC datasets used in the paper are also included (zip file).
\end{description}

\section*{Acknowledgments}\label{Acknowledgement}
The authors thank the editor, associate editor, and two referees, for their valuable comments that helped in improving the paper significantly. The authors acknowledge Advanced Research Computing at Virginia Tech for providing computational resources. The research was supported by National Science Foundation Grants CNS-1565314 and CNS-1838271 to Virginia Tech.

\section*{About the Authors}
\ \\[-4ex]
\indent\textbf{Li Xu} received his BS in statistics from Nanjing University in 2015 and MS in statistics from Virginia Tech in 2017. He is working toward the PhD degree in the Department of Statistics at Virginia Tech. His research interests include computer experiments, functional data analysis, spatial statistics, and machine learning.

\textbf{Thomas Lux} received the BS degree (cum laude) in computer science with minors in mathematics and physics from Roanoke College in 2016. He obtained his PhD degree in computer science at Virginia Polytechnic Institute and State University in 2020. His research interests include mathematical modeling, optimization, numerical analysis, and reinforcement learning for artificial intelligence.

\textbf{Tyler H. Chang} received his B.S. in Mathematics and Computer Science from Virginia Wesleyan College in 2016. He obtained his PhD degree in computer science at Virginia Polytechnic Institute and State University in 2020. His research interests include numerical analysis, algorithms, parallel computing, and data science.

\textbf{Bo Li} received the BS degree in automation from the Dalian University of Technology in 2006, the MS degree in control theory \& control engineering from the Dalian University of Technology in 2009, and the MS degree in computer science from Rochester Institute of Technology in 2012. He obtained his PhD degree in Computer Science from Virginia Tech in 2018. His research interests include power-aware computing in the high-performance computing domain and performance modeling of scientific parallel applications under DVFS, DCT, and DMT.

\textbf{Yili Hong} received the PhD in statistics from Iowa State University in 2009. He is Professor of Statistics at Virginia Tech. His research interests include machine learning and engineering applications, reliability analysis, and spatial statistics. He has more than 90 publications in venues such as \emph{Journal of the American Statistical Association}, \emph{Technometrics}, \emph{Journal of Quality Technology}, and \emph{Quality Engineering}. He is currently an associate editor for \emph{Technometrics} and \emph{Journal of Quality Technology}. He is an elected member of International Statistical Institute. He won the 2011 DuPont Young Professor Award, and the 2016 Frank Wilcoxon Prize in statistics.

\textbf{Layne T. Watson} (F '93) received the BA degree (magna cum laude) in psychology and mathematics from the University of Evansville, Indiana, in 1969, and the PhD degree in mathematics from the University of Michigan, Ann Arbor, in 1974. He has worked for USNAD Crane, Sandia National Laboratories, and General Motors Research Laboratories and served on the faculties of the University of Michigan, Michigan State University, and University of Notre Dame. He is currently a professor of computer science, mathematics, and aerospace and ocean engineering at Virginia Polytechnic Institute and State University. He serves as senior editor of \emph{Applied Mathematics and Computation}, and associate editor of \emph{Computational Optimization and Applications}, \emph{Evolutionary Optimization}, \emph{Engineering Computations}, and \emph{the International Journal of High Performance Computing Applications}. He is a fellow of the National Institute of Aerospace and the International Society of Intelligent Biological Medicine. He is a fellow of the IEEE. He has published well more than 300 refereed journal articles and 200 refereed conference papers. His research interests include fluid dynamics, solid mechanics, numerical analysis, optimization, parallel computation, mathematical software, image processing, and bioinformatics.

\textbf{Ali R. Butt} received the PhD degree in electrical and computer engineering from Purdue University. He is a professor of Computer Science and ECE (by courtesy) at Virginia Tech. He is a recipient of several awards such as an NSF CAREER Award. He is an alumni of the National Academy of Engineering's US Frontiers of Engineering (FOE) Symposium, US-Japan FOE, and National Academy of Science's AA Symposium on Sensor Science. Ali's research interests are in distributed computing systems, cloud/edge computing, file and storage systems, Internet of Things, I/O systems, and operating systems. At Virginia Tech he leads the Distributed Systems \& Storage Laboratory (DSSL).

\textbf{Danfeng (Daphne) Yao} is a Professor of Computer Science at Virginia Tech. She received her Ph.D. degree from Brown University, M.S. degrees from Princeton University and Indiana University, Bloomington, B.S. degree from Peking University in China. Her expertise is on software and system security, with a focus on detection and prediction accuracy. She was named an ACM Distinguished Scientist for her outstanding scientific contributions to cybersecurity. She received the NSF CAREER Award for her work on human-behavior driven malware detection and ARO Young Investigator Award for her semantic reasoning for mission-oriented security work. She has several Best Paper Awards and Best Poster Awards and currently leads multiple large-size multi-institution research projects.

\textbf{Kirk W. Cameron} received the PhD degree in computer science from Louisiana State University. He is professor of Computer Science at Virginia Tech and director of the stack@cs Center for Computer Systems. The central theme of his research is to improve performance and power efficiency in computer systems. Accolades for his research include U.S. National Science Foundation and U.S. Department of Energy Career Awards, IBM and AMD Faculty Awards, and a Distinguished Visiting fellowship from the U.K. Royal Academy of Engineering. He is member of the IEEE and a Distinguished Member of the ACM.


\begin{thebibliography}{}

\bibitem[\protect\citeauthoryear{{Bandler}, {Cheng}, {Dakroury}, {Mohamed},
  {Bakr}, {Madsen}, and {Sondergaard}}{{Bandler} et~al.}{2004}]{surrogatemodel}
{Bandler}, J.~W., Q.~S. {Cheng}, S.~A. {Dakroury}, A.~S. {Mohamed}, M.~H.
  {Bakr}, K.~{Madsen}, and J.~{Sondergaard} (2004).
\newblock Space mapping: the state of the art.
\newblock {\em IEEE Transactions on Microwave Theory and Techniques\/}~{\em
  52}, 337--361.

\bibitem[\protect\citeauthoryear{Ben-Ari and Steinberg}{Ben-Ari and
  Steinberg}{2007}]{BenAriSteinberg2007}
Ben-Ari, E.~N. and D.~M. Steinberg (2007).
\newblock Modeling data from computer experiments: An empirical comparison of
  {Kriging} with {MARS} and projection pursuit regression.
\newblock {\em Quality Engineering\/}~{\em 19}, 327--338.

\bibitem[\protect\citeauthoryear{Berry and Minser}{Berry and
  Minser}{1999}]{Berry:1999}
Berry, M.~W. and K.~S. Minser (1999).
\newblock Algorithm 798: High-dimensional interpolation using the modified
  {Shepard} method.
\newblock {\em ACM Transactions on Mathematical Software\/}~{\em 25}, 353--366.

\bibitem[\protect\citeauthoryear{Cameron, Anwar, Cheng, Xu, Li, Ananth,
  Bernard, Jearls, Lux, Hong, Watson, and Butt}{Cameron
  et~al.}{2019}]{Cameron-MOANA-2019}
Cameron, K.~W., A.~Anwar, Y.~Cheng, L.~Xu, B.~Li, U.~Ananth, J.~Bernard,
  C.~Jearls, T.~Lux, Y.~Hong, L.~T. Watson, and A.~R. Butt (2019).
\newblock {MOANA}: Modeling and analyzing {I/O} variability in parallel system
  experimental design.
\newblock {\em IEEE Transactions on Parallel and Distributed Systems\/}~{\em
  30}, 1843--1856.

\bibitem[\protect\citeauthoryear{Campolucci, Capperelli, Guarnieri, Piazza, and
  Uncini}{Campolucci et~al.}{1996}]{splineNN}
Campolucci, P., F.~Capperelli, S.~Guarnieri, F.~Piazza, and A.~Uncini (1996).
\newblock Neural networks with adaptive spline activation function.
\newblock In {\em Proceedings of 8th Mediterranean Electrotechnical Conference
  on Industrial Applications in Power Systems, Computer Science and
  Telecommunications (MELECON 96)}, Volume~3, pp.\  1442--1445.

\bibitem[\protect\citeauthoryear{Chen, Tsui, Barton, and Meckesheimer}{Chen
  et~al.}{2006}]{compexpsummary}
Chen, V.~C., K.-L. Tsui, R.~R. Barton, and M.~Meckesheimer (2006).
\newblock A review on design, modeling and applications of computer
  experiments.
\newblock {\em IIE Transactions\/}~{\em 38}, 273--291.

\bibitem[\protect\citeauthoryear{Chipman, George, and McCulloch}{Chipman
  et~al.}{1998}]{CGM98}
Chipman, H.~A., E.~I. George, and R.~E. McCulloch (1998).
\newblock Bayesian {CART} model search.
\newblock {\em Journal of the American Statistical Association\/}~{\em 93},
  935--948.

\bibitem[\protect\citeauthoryear{Chipman, George, and McCulloch}{Chipman
  et~al.}{2002}]{Chipman2002}
Chipman, H.~A., E.~I. George, and R.~E. McCulloch (2002).
\newblock Bayesian treed models.
\newblock {\em Machine Learning\/}~{\em 48}, 299--320.

\bibitem[\protect\citeauthoryear{Chipman, George, and McCulloch}{Chipman
  et~al.}{2010}]{bartanas}
Chipman, H.~A., E.~I. George, and R.~E. McCulloch (2010).
\newblock {BART}: {Bayesian} additive regression trees.
\newblock {\em The Annals of Applied Statistics\/}~{\em 4}, 266--298.

\bibitem[\protect\citeauthoryear{Choi, Shi, and Wang}{Choi
  et~al.}{2011}]{ChoiShiWang2011}
Choi, T., J.~Shi, and B.~Wang (2011).
\newblock A {Gaussian} process regression approach to a single-index model.
\newblock {\em Journal of Nonparametric Statistics\/}~{\em 23}, 21--36.

\bibitem[\protect\citeauthoryear{Deng, Hung, and Lin}{Deng
  et~al.}{2015}]{DengHungLin2015}
Deng, X., Y.~Hung, and C.~D. Lin (2015).
\newblock Design for computer experiments with qualitative and quantitative
  factors.
\newblock {\em Statistica Sinica\/}~{\em 25}, 1567--1581.

\bibitem[\protect\citeauthoryear{Friedman}{Friedman}{1991}]{friedman1991}
Friedman, J.~H. (1991).
\newblock Multivariate adaptive regression splines.
\newblock {\em The Annals of Statistics\/}~{\em 19}, 1--67.

\bibitem[\protect\citeauthoryear{Gordon and Wixom}{Gordon and
  Wixom}{1978}]{Gordon:1978}
Gordon, W.~J. and J.~A. Wixom (1978).
\newblock Shepard's method of ``metric interpolation'' to bivariate and
  multivariate interpolation.
\newblock {\em Mathematics of Computation\/}~{\em 32}, 253--264.

\bibitem[\protect\citeauthoryear{Gramacy and Lee}{Gramacy and
  Lee}{2008}]{tgpjasa}
Gramacy, R.~B. and H.~K.~H. Lee (2008).
\newblock Bayesian treed {Gaussian} process models with an application to
  computer modeling.
\newblock {\em Journal of the American Statistical Association\/}~{\em 103},
  1119--1130.

\bibitem[\protect\citeauthoryear{Gramacy and Lian}{Gramacy and
  Lian}{2012}]{GramacyLian2012}
Gramacy, R.~B. and H.~Lian (2012).
\newblock {Gaussian} process single-index models as emulators for computer
  experiments.
\newblock {\em Technometrics\/}~{\em 54}, 30--41.

\bibitem[\protect\citeauthoryear{Hastie, Tibshirani, and Friedman}{Hastie
  et~al.}{2009}]{statisticallearning}
Hastie, T., R.~Tibshirani, and J.~Friedman (2009).
\newblock {\em The Elements of Statistical Learning}.
\newblock New York: Springer.

\bibitem[\protect\citeauthoryear{He, Watson, and Sosonkina}{He
  et~al.}{2009}]{vtdirect95}
He, J., L.~T. Watson, and M.~Sosonkina (2009).
\newblock Algorithm 897: {VTDIRECT95}: Serial and parallel codes for the global
  optimization algorithm direct.
\newblock ~{\em 36, article no. 17}.

\bibitem[\protect\citeauthoryear{Hestenes}{Hestenes}{1969}]{Hestenes1969}
Hestenes, M.~R. (1969).
\newblock Multiplier and gradient methods.
\newblock {\em Journal of Optimization Theory and Applications\/}~{\em 4},
  303--320.

\bibitem[\protect\citeauthoryear{Hung, Joseph, and Melkote}{Hung
  et~al.}{2015}]{HungJosephMelkote2015}
Hung, Y., V.~R. Joseph, and S.~N. Melkote (2015).
\newblock Analysis of computer experiments with functional response.
\newblock {\em Technometrics\/}~{\em 57}, 33--54.

\bibitem[\protect\citeauthoryear{Joseph}{Joseph}{2016}]{Joseph2016}
Joseph, V.~R. (2016).
\newblock Space-filling designs for computer experiments: A review.
\newblock {\em Quality Engineering\/}~{\em 28}, 28--35.

\bibitem[\protect\citeauthoryear{Kim and Gu}{Kim and Gu}{2004}]{RSSB}
Kim, Y.-J. and C.~Gu (2004).
\newblock Smoothing spline {Gaussian} regression: more scalable computation via
  efficient approximation.
\newblock {\em Journal of the Royal Statistical Society: Series B (Statistical
  Methodology)\/}~{\em 66}, 337--356.

\bibitem[\protect\citeauthoryear{Lux, Watson, Chang, Bernard, Li, Yu, Xu, Back,
  Butt, Cameron, Hong, and Yao}{Lux et~al.}{2018}]{lux2018nonparametric}
Lux, T.~C., L.~T. Watson, T.~H. Chang, J.~Bernard, B.~Li, X.~Yu, L.~Xu,
  G.~Back, A.~R. Butt, K.~W. Cameron, Y.~Hong, and D.~Yao (2018).
\newblock Nonparametric distribution models for predicting and managing
  computational performance variability.
\newblock In {\em SoutheastCon 2018}, pp.\  1--7. IEEE.

\bibitem[\protect\citeauthoryear{Myers, Montgomery, and {Anderson-Cook}}{Myers
  et~al.}{2016}]{MyersMontgomeryAnderson-Cook2016}
Myers, R.~H., D.~C. Montgomery, and C.~M. {Anderson-Cook} (2016).
\newblock {\em Response Surface Methodology: Process and Product Optimization
  Using Designed Experiments\/} (4th ed.).
\newblock Hoboken, NJ: John Wiley \& Sons.

\bibitem[\protect\citeauthoryear{Qian, Wu, and Wu}{Qian
  et~al.}{2008}]{QianWuWu2008}
Qian, P. Z.~G., H.~Wu, and C.~F.~J. Wu (2008).
\newblock Gaussian process models for computer experiments with qualitative and
  quantitative factors.
\newblock {\em Technometrics\/}~{\em 50}, 383--396.

\bibitem[\protect\citeauthoryear{Renka}{Renka}{1988}]{Renka1}
Renka, R.~J. (1988).
\newblock Algorithm 660: {QSHEP2D}: Quadratic {Shepard} method for bivariate
  interpolation of scattered data.
\newblock {\em ACM Transactions on Mathematical Software\/}~{\em 14}, 149--150.

\bibitem[\protect\citeauthoryear{Santner, Williams, and Notz}{Santner
  et~al.}{2010}]{santner2010design}
Santner, T., B.~Williams, and W.~Notz (2010).
\newblock {\em The Design and Analysis of Computer Experiments}.
\newblock New York: Springer.

\bibitem[\protect\citeauthoryear{Shepard}{Shepard}{1968}]{Shepard}
Shepard, D. (1968).
\newblock A two-dimensional interpolation function for irregularly-spaced data.
\newblock In {\em Proceedings of the 1968 23rd ACM National Conference}, ACM
  '68, New York, pp.\  517--524. ACM.

\bibitem[\protect\citeauthoryear{Taddy, Gramacy, and Polson}{Taddy
  et~al.}{2011}]{dynatree}
Taddy, M.~A., R.~B. Gramacy, and N.~G. Polson (2011).
\newblock Dynamic trees for learning and design.
\newblock {\em Journal of the American Statistical Association\/}~{\em 106},
  109--123.

\bibitem[\protect\citeauthoryear{Thacker, Zhang, Watson, Birch, Iyer, and
  Berry}{Thacker et~al.}{2010}]{Thacker}
Thacker, W.~I., J.~Zhang, L.~T. Watson, J.~B. Birch, M.~A. Iyer, and M.~W.
  Berry (2010).
\newblock Algorithm 905: {SHEPPACK}: Modified {Shepard} algorithm for
  interpolation of scattered multivariate data.
\newblock {\em ACM Transactions on Mathematical Software\/}~{\em 37},
  34:1--34:20.

\bibitem[\protect\citeauthoryear{Xu, Wang, Lux, Chang, Bernard, Li, Hong,
  Cameron, and Watson}{Xu et~al.}{2020}]{Xuetal-JPDC-2020}
Xu, L., Y.~Wang, T.~Lux, T.~Chang, J.~Bernard, B.~Li, Y.~Hong, K.~Cameron, and
  L.~Watson (2020).
\newblock Modeling {I/O} performance variability in high-performance computing
  systems using mixture distributions.
\newblock {\em Journal of Parallel and Distributed Computing\/}~{\em 139},
  87--98.

\bibitem[\protect\citeauthoryear{Zhang and Notz}{Zhang and
  Notz}{2014}]{ZhangNotz2014}
Zhang, Y. and W.~I. Notz (2014).
\newblock Computer experiments with qualitative and quantitative variables: A
  review and reexamination.
\newblock {\em Quality Engineering\/}~{\em 27}, 2--13.

\bibitem[\protect\citeauthoryear{Zhou, Qian, and Zhou}{Zhou
  et~al.}{2011}]{ZhouQianZhou2011}
Zhou, Q., P.~Z.~G. Qian, and S.~Zhou (2011).
\newblock A simple approach to emulation for computer models with qualitative
  and quantitative factors.
\newblock {\em Technometrics\/}~{\em 53\/}(3), 266--273.

\end{thebibliography}

\end{document}